\documentclass[11pt]{article}
\pdfoutput=1

\usepackage{setspace}
\usepackage[T1]{fontenc}                
\usepackage{microtype} 				    
\usepackage{anyfontsize}                
\usepackage{amssymb,amsmath,amscd}
\usepackage[UKenglish]{babel}
\usepackage{titlesec}
\usepackage{graphicx,xcolor,xparse}
\usepackage{enumitem}
\usepackage{multirow}
\usepackage[absolute]{textpos}
\usepackage[margin=1.5em,labelfont=bf]{caption}
\usepackage{cite}
\usepackage{mathrsfs}                           
\usepackage[hang,flushmargin]{footmisc}         
\usepackage{orcidlink}                          

\usepackage{hyperref}                           
\hypersetup{
	colorlinks=true,
	linkcolor=dark-blue,
	citecolor=dark-red,
	urlcolor=dark-green,
	linktoc=all
}

\usepackage[vcentermath]{youngtab}
\usepackage{tikz} 
\usepackage{tikz-cd} 
\usetikzlibrary{matrix,calc} 
\usetikzlibrary{decorations.markings,shapes.geometric,shapes.misc} 
\usetikzlibrary{decorations.pathreplacing,positioning} 
\usetikzlibrary{arrows}

\usepackage{geometry} 					
\numberwithin{equation}{section} 

\titleformat{\section}[block]{\Large\bfseries\centering}{\thesection}{1em}{} 
\titleformat{\subsection}[block]{\bfseries}{\thesubsection}{1em}{} 

\titlespacing*{\section}{0pt}{1em}{1em}
\titlespacing*{\subsection}{0pt}{0.75em}{0.75em}

\addto\captionsUKenglish{
	\renewcommand{\contentsname}%
	{Table of Contents}%
}

\newcommand\TBstrut{\rule[-0.9ex]{0pt}{3.5ex}}   

\definecolor{dark-gray}{gray}{0.20}
\definecolor{gray}{gray}{0.30}
\definecolor{light-gray}{gray}{0.80}
\definecolor{dark-red}{rgb}{0.7,0,0}
\definecolor{dark-green}{rgb}{0.1,0.4,0}
\definecolor{dark-blue}{rgb}{0.3,0.3,0.7}
\definecolor{light-blue}{rgb}{0.8,0.8,1}
\definecolor{cardinal}{rgb}{0.6,0,0}
\definecolor{darkgreen}{rgb}{0,0.5,0}
\definecolor{golden}{rgb}{0.92, 0.7, 0}
\definecolor{midnight}{rgb}{0, 0, 0.5}
\definecolor{darkblue}{rgb}{0.2, 0, 0.8}
\definecolor{forestgreen}{rgb}{0.13, 0.55, 0.13}
\definecolor{darkred}{rgb}{0.7,0,0}

\def\mop#1{\mathop{\rm #1}\nolimits}

\def\ii{{\rm i}}

\def\AdS{\mop{AdS}}

\newcommand{\ds}{{\rm d}s}
\newcommand{\dd}{{\rm d}}
\newcommand{\DD}{{\rm D}}
\newcommand{\e}{\mathrm{e}}

\newcommand{\dvol}{{\rm vol}}

\renewcommand{\gcd}{\mathrm{gcd}}

\newcommand{\spindle}{{\includegraphics[height=8pt]{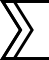}}}
\newcommand{\sspindle}{\hspace{1pt}{\includegraphics[height=5pt]{mySpindle.png}}}

\newcommand{\disc}{\mathbb{D}}

\newcommand{\f}[2]{\frac{#1}{#2}}

\newcommand{\nn}{\nonumber}
\newcommand{\wti}[1]{\widetilde{#1}}

\def\overleftrightarrow#1{\vbox{\ialign{##\crcr
			$\leftrightarrow$\crcr\noalign{\kern-0pt\nointerlineskip}
			$\hfil\displaystyle{#1}\hfil$\crcr}}}

\newcommand\bC{\mathbf{C}}

\newcommand\bP{\mathbf{P}}
\newcommand\bQ{\mathbf{Q}}
\newcommand\bR{\mathbf{R}}

\newcommand\bZ{\mathbf{Z}}

\newcommand\cA{\mathcal{A}}

\newcommand\cC{\mathcal{C}}
\newcommand\cD{\mathcal{D}}

\newcommand\cF{\mathcal{F}}

\newcommand\cI{\mathcal{I}}

\newcommand\cL{\mathcal{L}}
\newcommand\cM{\mathcal{M}}
\newcommand\cN{\mathcal{N}}

\newcommand\cS{\mathcal{S}}

\newcommand\fg{\mathfrak{g}}

\newcommand\fr{\mathfrak{r}}

\newcommand\SL{\mathrm{SL}}

\newcommand\SO{\mathrm{SO}}

\newcommand\UU{\mathrm{U}}
\newcommand\SU{\mathrm{SU}}

\newcommand\su{\mathfrak{su}}

\newcommand\uu{\mathfrak{u}}

\title{~\vspace{10mm}\\
	\fontsize{20pt}{23pt}\selectfont\textbf{On the Class $\cS$ Origin of Spindle Solutions}\vspace{10mm}}

\author{\Large{Pieter Bomans\textsuperscript{\orcidlink{0000-0002-0907-9830}} and Christopher Couzens\textsuperscript{\orcidlink{0000-0001-9659-8550}}}\\[10mm]
	\large Mathematical Institute, University of Oxford\\
	\normalsize Andrew Wiles Building, Radcliffe Observatory Quarter\\
	\normalsize Woodstock Road, Oxford, OX2 6GG, U.K.\\[5mm]
	\texttt{\normalsize\href{mailto:pieter.bomans@maths.ox.ac.uk}{pieter.bomans@maths.ox.ac.uk}, \href{mailto:christopher.couzens@maths.ox.ac.uk}{christopher.couzens@maths.ox.ac.uk}}\\
}

\date{}

\begin{document}
	
	\pagestyle{empty}
	
	\maketitle
	\thispagestyle{empty}

	\vspace{\stretch{1}}
	
	\begin{abstract}
		\noindent We analyse the backreacted geometry corresponding to a stack of M5-branes wrapped on a spindle, with a view towards precision tests of the dual $\cN=1$ superconformal field theory. We carefully study the singular loci of the uplifted geometry and show that these correspond to $\bC^3/\bZ_n$ conical singularities. Therefore, these solutions present one of the first explicit realisations of honest locally $\cN=1$ preserving punctures in class $\cS$. Additionally we study the symmetries and anomalies of the dual field theory through anomaly inflow and compute a variety of holographic observables including dimensions of BPS operators. This work paves the way for advancements in the study and identification of the precise dual field theories.
	\end{abstract}
	
	\vspace{\stretch{3}}
	
	\newpage

	{ 
		\hypersetup{linkcolor=black}
		\setcounter{tocdepth}{2}
		\setlength{\parskip}{0.4em}
		\tableofcontents
	}
	
	\setcounter{page}{0}
	
	\clearpage
	\pagestyle{plain}
	
	\section{Introduction and summary}      %
	\label{sec:intro}                       %
	
	Placing a supersymmetric field theory on a curved background generically completely breaks supersymmetry. However, one can preserve supersymmetry by coupling the theory to a collection of (rigid supersymmetry) background fields \cite{Festuccia:2011ws}. Perhaps the most well known example of this mechanism is the (partial) topological twist \cite{Witten:1988ze,Witten:1994ev,Johansen:1994aw}. Schematically, the Killing spinor equation takes the form 
	\begin{equation}
		\delta_\epsilon\psi_\mu \sim (\partial_\mu +\f14 \omega_\mu \cdot \gamma + A_\mu)\epsilon = 0\,,
	\end{equation}
	and by turning on a background gauge field $A_\mu \propto \omega_\mu$, which compensates for the curvature of spacetime, we can solve the Killing spinor equation with a constant spinor $\epsilon$, i.e. $\partial_\mu \epsilon=0$. In string or M-theory this scenario is realised by wrapping M- or D-branes on calibrated cycles where supersymmetry is preserved by turning on the gauge connection on the normal bundle of the cycle such that it cancels the spin connection of the cycle. This is not the only mechanism however.
	
	The focus of this note lies in compactifications of the six-dimensional $\cN=(2,0)$ theory on a complex curve $\cC$. In this case the curved manifold is of the form $\cM_{4}\times \cC$, where $\cC$ is a complex curve of genus $g$ possibly decorated with punctures. The resulting four-dimensional theories are called theories of class $\cS$ \cite{Gaiotto:2009gz} and preserve either $\cN=1$ or $\cN=2$ supersymmetry, depending on the choice of background flux. Even though the resulting theories are often non-Lagrangian, the latter case is under relatively good control and a variety of tools to study them are available, such as the associated Hitchin system \cite{Gaiotto:2009hg} or vertex operator algebras \cite{Beem:2013sza,Beem:2014rza}. Moreover, it is believed that they are classified by the type of 6d $(2,0)$ theory, the choice of Riemann surface and the additional local data specifying the punctures. 
	
	The landscape of $\cN=1$ class $\cS$ theories on the other hand remains largely unexplored. Although a variety of punctures have been discussed in an $\cN=1$ setting, most of them, such as the $(p,q)$ punctures discussed in \cite{Bah:2015fwa,Bah:2018gwc,Bah:2019jts}, locally preserve $\cN=2$ supersymmetry but are glued together so that globally only $\cN=1$ supersymmetry is preserved. Alternatively, in \cite{Bah:2013wda}, $\cN=1$ punctures were discussed in a probe approximation but the backreacted geometry remained unknown. In the $\cN=2$ setting the allowed punctures can be understood as allowed singular boundary conditions of the associated Hitchin system. Even though a generalised Hitchin system is believed to govern the allowed punctures in $\cN=1$ theories \cite{Xie:2013gma,Xie:2013rsa,Xie:2014yya} and a variety of special cases has been considered (see \cite{Benini:2009mz,Tachikawa:2011ea,Beem:2012yn,Gadde:2013fma,Maruyoshi:2013hja,Bonelli:2013pva,Bah:2012dg,Bah:2011vv,Bah:2013aha,Agarwal:2014rua,Giacomelli:2014rna} for a necessarily incomplete set of references), a satisfactory mathematical and physical understanding of this problem has yet to be uncovered.
	
	A fruitful approach to make progress in this direction is provided by holography. In M-theory these solutions can be obtained by wrapping a stack of M5-branes on the complex curve $\cC$. Punctures in turn can be described by additional flavour branes intersecting the stack of branes at a point on $\cC$. From a UV point of view, this setup can be described by a stack of M5 branes wrapping a complex curve inside a Calabi-Yau three-fold, $X$. The local geometry of $X$ is given by two line bundles over $\cC$, $\cL_1 \oplus \cL_2 \hookrightarrow X \rightarrow \cC\,$, encoding the topology of $X$. The condition to preserve supersymmetry then reduces to the requirement that the determinant line bundle is the canonical bundle over the curve, i.e. $\cL_1 \otimes \cL_2 = K_\cC$.\footnote{When $q_1$ or $q_2$ vanishes the three-fold geometry simplifies to the direct product $X = \bC \times T^*\cC$ and consequently $\cN=2$ supersymmetry is preserved.} In terms of the Chern numbers $p_1=c_1(\cL_1)$ and $p_2=c_1(\cL_2)$ this condition takes the form $p_1+p_2 = \chi(\cC)$ where $\chi(\cC)$ is the Euler characteristic of the curve.
	
	From a low energy point of view, these brane setups can be represented by backreacted $\AdS_5\times \cM_6$ solutions of eleven-dimensional supergravity where $\cM_6$ is an $S^4$ fibration over $\cC$. Such solutions were first considered in \cite{Maldacena:2000mw} and later extended in \cite{Bah:2011vv,Bah:2012dg} to include more general fluxes. Furthermore, in the case of smooth Riemann surfaces, it was shown in \cite{Anderson:2011cz,Bobev:2020jlb} that the metric on the curve flows towards its constant curvature form along the RG flow. This is in line with observations in ($\cN=2$) class $\cS$ theories where the four-dimensional theory is seen to be independent of all the K\"ahler deformations of the curve.\footnote{In addition, complex structure deformations as well as the addition of non-trivial flat gauge fields correspond to exactly marginal deformations of the dual SCFT.}
	
	The class $\cS$ theories reviewed above can include additional local contributions at marked points on the curve. The local geometry around regular punctures in the $\cN=2$ class $\cS$ was studied in \cite{Gaiotto:2009gz} where they were described as a collection of singular solutions to the Toda equation. For a specific class of punctures it was shown in \cite{Bobev:2018ugk} that such singular solutions correspond to conical defects on the curve. When considering curves of genus $g\geq 1$ or $g=0$ with at least three punctures all the above arguments go through identically upon converting the usual bundles and characteristic classes into orbibundles and orbifold characteristic classes. Indeed, in this case supersymmetry is still preserved by the same topological twist and the IR metric on the curve remains the constant curvature one.
	
	In this work, we consider a novel situation in which supersymmetry is not preserved through the usual topological twist. Indeed, a natural way to circumvent the results of \cite{Anderson:2011cz} is to consider curves which do not possess a constant curvature metric. Such cases are provided by the so-called bad orbifolds \cite{thurston2022geometry} of which there exist only two cases. The spindle or the teardrop, i.e. a sphere with one or two conical defects, denoted by $\spindle$,\footnote{More precisely, the integers $n_\pm$ parameterising the deficit angles $2\pi\left(1-\f{1}{n_\pm}\right)$ have to be coprime.} or a disc, $\disc$, with one conical defect at its centre. We focus on the spindle for the remainder of this note.  
	
	The relevant spindle solutions were constructed in maximal $\SO(6)$ gauged supergravity in \cite{Ferrero:2021wvk} while the disc solutions were constructed in \cite{Bah:2021mzw,Bah:2021hei} with a variety of generalisations considered in \cite{Couzens:2021tnv,Suh:2021ifj,Couzens:2022yjl,Bah:2022yjf,Bomans:2023ouw,Couzens:2023kyf,Macpherson:2024frt,Merrikin:2024bmv}. For the disc solutions it is possible to preserve $\cN=2$ supersymmetry and in this case the dual SCFTs are of Argyres-Douglas type \cite{Argyres:1995jj,Xie:2012hs}.\footnote{Note that for the disc, the mechanism for preserving supersymmetry is different than that for the spindle. In this case the canonical bundle of $X$ is not trivial and hence it is not Calabi-Yau. This situation is somewhat closer to the anti-twist scenario discussed in \cite{Ferrero:2021etw}. Since the focus of this note is on the spindle we do not discuss this case in detail.} The boundary of the disc was understood to correspond to the irregular puncture, while the regular puncture originates from the conical defect at the centre of the disc. For the spindle on the other hand, there are no consistent solutions preserving $\cN=2$ supersymmetry and one is forced to consider $\cN=1$ supersymmetric solutions. In the case of the two-sphere with up to two punctures it is understood that the dual theories do not flow to an SCFT but instead to a topological sigma-model \cite{Assel:2016lad}. 
	
	In both situations above, the complex curve does not allow for a constant curvature metric, and consequently supersymmetry is realised in a distinct way. When wrapping a stack of M5-branes on the spindle, the UV picture is identical to the usual topological twist, however, the IR picture is rather different. Indeed, the constraint $p_1 + p_2 = \chi\left(\,\spindle\,\right)$ is still satisfied, but locally the spin connection is not cancelled by the R-symmetry gauge fields, nor are the spinors constant. In other words, supersymmetry is preserved by a topologically topological twist.
	
	Given this novel mechanism for preserving supersymmetry a natural question is: \textit{What is the dual field theory?}. In this note we aim to make progress on this open question by carefully investigating the eleven-dimensional solutions corresponding to the spindle solutions, analysing their symmetries and computing a slew of holographic observables.

	\paragraph{Summary of the results}~
	
	In order to extract more detailed information about the dual SCFT it proves useful to carefully uplift the seven-dimensional solution to eleven-dimensional supergravity. Doing so allows us to scrutinise the internal space and give a detailed description of the geometry. Unlike the analogous M2-, D3- solutions \cite{Ferrero:2020laf,PhysRevD.104.046007,Ferrero:2021ovq,Hosseini:2021fge,Boido:2021szx,Couzens:2021rlk,Faedo:2021nub,Couzens:2021cpk,Giri:2021xta,Boido:2022mbe,Couzens:2022yiv,Hristov:2023rel} the M5-brane spindle solutions remain singular after uplifting them to eleven dimensions.\footnote{The analogous D4/D8-system also admits singularities in the uplift.}
	
	We carefully demonstrate that each orbifold singularity of the spindle results in a pair of orbifold singularities in the uplifted geometries, locally taking the form $\AdS_5 \times \bC^3/\bZ_{n_\pm}$ where the precise form of the orbifold action is determined by the orbibundle data of the seven-dimensional solution. Similar to the $\cN=2$ class $\cS$ story, such punctures generate additional flavour symmetries in the dual SCFT. Indeed, this can be seen from expanding the M-theory three-form potential, or equivalently the four-form flux, on the internal cycles of the manifold. Analysing the crepant resolutions of such orbifold singularities, following Ito-Reid \cite{Ito-Reid}, we show that there are additional localised compact cycles dual to the exceptional divisors of the blow-up that give rise to vector fields for the additional global symmetries. We carefully analyse the resulting flavour symmetries for a variety of orbifold actions and describe the precise form of the possibly non-abelian flavour symmetry as well as the flavour anomalies $k_{\alpha\beta\gamma}$. For example, for the two infinite classes of punctures $\bC^3/\bZ_{n}$ with $n=2k+1$ or $n=2k+2$ and orbifold action generated by $\f1n(1,1,n-2)$ we find that the associated flavour symmetry of the dual SCFT is given by $\SU(k)^2$, where each of the pair of orbifold singularities of the uplifted geometry gives rise to one factor. This gives the first explicit holographic description of honestly, locally, $\cN=1$ preserving punctures and opens the way for further explorations.
	
	In addition to finding the global symmetry, we carefully study the anomalies of the four-dimensional SCFT through anomaly inflow by integrating the twelve-form characteristic class $\cI_{12}$ \cite{Harvey:1998bx} over the internal manifold $\cM_6$, including the effects from the two-cycles resulting from the blow-up. Doing so, we show that in contrast to situation on the disc, where one of the flavour symmetries is broken through a St\"uckelberg mechanism, there is no obstruction to equivariantly complete the four-form flux $G_4$ to a $\UU(1)^3$ equivariant form. Therefore, this firmly establishes that the dual SCFT does in fact inherit the full $\UU(1)^3$ global symmetry originating from the isometries of the internal manifold.\footnote{The work \cite{Bomans:2023ouw} gives an alternative viewpoint for the breaking of the flavour symmetry. There it was shown that the disc solutions can be generalised to include a scalar field which explicitly breaks the $\UU(1)$ in the uplift. Turning on such symmetry breaking scalar fields is not possible for the spindle solution, and thus one expects that these are genuine flavour symmetries of the dual field theory.} Finally, we compute the conformal dimensions of a variety of chiral primaries corresponding to M2-branes wrapping calibrated cycles as well as study the allowed probe M5 branes allowing us to probe a rich spectrum of information of the dual SCFT.

	\vspace{-8pt}\paragraph{Outlook}~
	
	In this work we have carefully analysed the local geometry describing the singular loci of the uplifted solutions corresponding to M5-branes wrapped on a spindle. We found that these singular loci describe locally $\cN=1$ punctures and explained how to extract a variety of observables of the dual SCFTs from such singularities.
	
	The type of singularities we study is far from the most general class. In order to preserve $\cN=1$ supersymmetry such singularities are restricted to be locally $\bC^3/\Gamma$, where $\Gamma$ is a finite subgroup of $\SU(3)$. Such subgroups have been classified \cite{finitesubgroups1,finitesubgroups2,finitesubgroups3,finitesubgroups4} and include more options then the abelian orbifold with $\Gamma=\bZ_n$ that arise in this work. There are more general abelian orbifolds with $\Gamma=\bZ_m\times \bZ_n$ as well as a variety of non-abelian finite subgroups $\Gamma \subset \SU(3)$. It would be interesting to investigate whether we can find generalisations of the solutions described in this work along the lines of \cite{Gaiotto:2009gz,Couzens:2022yjl,Bah:2022yjf,Couzens:2023kyf}. To make progress in this direction it may be beneficial to embed the solutions found here in the framework of \cite{Bah:2015fwa} and look for a generalisation of the $\cN=2$ electrostatics picture allowing one to generalise the solutions to the relevant Monge-Amp\'ere equation. 
	
	Further, having described such general punctures on the spindle a natural question is whether we can construct solutions describing M5-branes on higher genus Riemann surfaces and include these more general punctures. Similar to the $\cN=2$ case, finding global solutions with a variety of punctures is prohibitively hard. However, using the newly developed tools of equivariant localisation in supergravity \cite{BenettiGenolini:2023kxp,BenettiGenolini:2023yfe,BenettiGenolini:2023ndb,BenettiGenolini:2024kyy,Suh:2024asy,BCtoappear2} it should nonetheless be possible to extract a wealth of information from a careful local analysis of the relevant singularities.
	
	Finally, in this note we studied in detail the internal geometry corresponding to the M5 spindle solutions. We shed new light on a variety of observables and clarified various aspects of the dual SCFTs. However, it would be very interesting to exactly determine the dual SCFT with said properties. The variety of punctures described holographically could have a dual description as $\cN=1$ quiver tails which could be added to SCFTs such as BBBW and its generalisations providing a new larger set of $\cN=1$ class $\cS$ SCFTs. We hope to come back to these question in the future.

	\vspace{-8pt}\paragraph{Structure of this paper}~
	
	The remainder of this paper is organised as follows. In Section \ref{sec:spindle}, we revisit the spindle solutions of seven-dimensional gauged supergravity and carefully discuss the defining data of the relevant orbibundles. We proceed in Section \ref{sec:uplift} by presenting the explicit uplift of the solution to eleven-dimensions and analyse the singular loci of the eleven-dimensional geometry. Having done so, we continue our analysis in Section \ref{sec:observables} and \ref{sec:anomalies} where we study respectively the holographic observables originating from wrapped branes and the global symmetries and anomalies of the dual SCFT . In Appendices \ref{app:upliftformulae} and \ref{app:GSMW} we summarise the uplift formulae and describe the eleven-dimensional solution in canonical $\cN=1$ coordinates. Finally, in Appendix \ref{app:BBBW} we show how in a set of specific (singular) limits of the solutions presented in this note one can recover the BBBW solutions corresponding to M5 branes wrapping a smooth (higher-genus) Riemann surface. This allows us to better contrast the new ingredients of the spindle field theories against the more established BBBW solution and dual field theory.

	\section{Seven-dimensional spindle solution} %
	\label{sec:spindle}                                      %
	
	The main character of this work is the spindle solution of seven-dimensional gauged supergravity. This section serves as an introduction to the solution, in particular the conventions we will employ in this work. This background is a solution to the seven-dimensional $\UU(1)^2\subset \SO(5)$ gauged supergravity BPS equations and equations of motion \cite{Pernici:1985ju}.\footnote{We use the same conventions as in \cite{Bomans:2023ouw}.} The relevant solution was described in \cite{Ferrero:2021wvk,Ferrero:2021etw} and is given as follows. The metric, gauge fields and scalar are
	\begin{equation}\label{eq:7dM5sol}
		\begin{aligned}
			\ds^2_7 &=\, \Big(w H(w)\Big)^{1/5}\left[ \ds^2_{\AdS_5} + \f{w}{f(w)}\dd w^2+\f{f(w)}{4H(w)} \fr_z^2\dd z^2\right]\,,\\
			A^{(i)} &=\, -\f{w^2}{h_i(w)}\fr_z\,\dd z\, ,\qquad\qquad X^{(i)}(w)=\frac{\left(w H(w)\right)^{2/5}}{h_i(w)}\,,
		\end{aligned}
	\end{equation}
	where $i=1,2$ and $\dd s^2_{\AdS_5}$ is the metric on $\AdS_5$ with unit radius, satisfying $R_{\mu\nu}=-4 g_{\mu\nu}$. We explicitly include the constant $\fr_z$ parameterizing the radius of the $z$ circle such that the coordinate $z$ always has period $2\pi$. The functions $f$, $H$ and $h_i$ take the form
	\begin{equation}\label{eq:FHfuncs}
		h_{i}(w)= q_i + w^2 \,,\qquad 
		H(w) = h_1(w)h_2(w) \,,\qquad 
		f(w) = H(w) - 4 w^3 \,.
	\end{equation}
	\begin{figure}[!ht]
		\centering
		\includegraphics[width=0.5\textwidth]{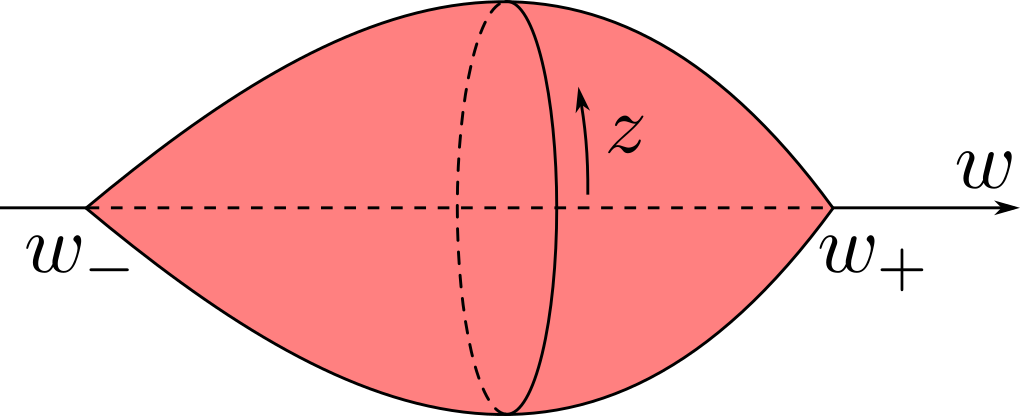}
		\caption{A pictorial representation of a spindle.}
		\label{fig:spindle}
	\end{figure}

	In this work we are interested in spindle solutions, i.e. compact global completions of the local solution above which have the topology of a sphere with one or two conical defects at the poles.\footnote{The local solution also contains compact solutions with disc topology and non-compact co-dimension 2 defect solutions \cite{Gutperle:2022pgw,Gutperle:2023yrd,Capuozzo:2023fll} as different global completions. Much of the analysis of the probe branes and singularity structure of the solutions here goes over for the $\mathcal{N}=1$ non-compact solutions of the latter references.  } For this reason we restrict ourselves to the portion of $(q_1,q_2)$ parameter space with $q_1q_2<0$ and where the function $f$ has four distinct real roots and restrict the range of $w \in [w_-,w_+]$ to lie between the middle two roots where the function $f$ is positive. One can slightly enlarge the truncation and for each of the two $\UU(1)$'s add a charged scalar which in the dual SCFT breaks the relevant $\UU(1)$ global symmetry. However, as shown in \cite{Bomans:2023ouw} such scalars cannot be turned on while preserving the sphere topology and we will consider them further in this work. 
	
	We denote the two-dimensional spindle metric by
	\begin{equation}
		\ds^2_{\sspindle} = \f{w}{f(w)}\dd w^2 + \f{f(w)}{4H(w)} \fr_z^2\dd z^2\,.
	\end{equation}
	Near one of the poles, i.e. near a root of $f(w)$ the two-dimensional metric takes the form
	\begin{equation}
		\ds^2_{\sspindle} \rightarrow  \f{4w_\pm}{|f'(w_\pm)|}\left(\dd r^2 + \f{|f'(w_\pm)|^2 \fr_z^2}{64 w_\pm^4}  \, r^2\,\dd z^2\right) \,, \qquad w\rightarrow w_\pm\,,
	\end{equation}
	where we changed the radial coordinate to $r^2= \mp (w-w_\pm)$ at $w \simeq w_\pm$ respectively. Locally, around each root we therefore recover the metric of an $\bR^2/\bZ_{n_\pm}$ orbifold, provided that the radius $\fr_z$ satisfies,
	\begin{equation}\label{eq:orbsing2d}
		\fr_z = \f{8w_\pm^2}{n_\pm|f'(w_\pm)|}\,.
	\end{equation}
	The fluxes through the spindle are given by
	\begin{equation}
		Q_i \equiv \f1{2\pi}\int_{\sspindle} F^{(i)} = - \fr_z\,\f{w^2}{h_i(w)}\Big|^{w=w_+}_{w=w_-} = \f{p_i}{n_+n_-}\,, \qquad\qquad p_i \in \bZ\,,
	\end{equation}
	in line with the quantisation conditions for $\UU(1)$-orbibundles over the spindle \cite{Bobev:2019ore,Ferrero:2021etw}. The orbifold Euler characteristic of the spindle is given by
	\begin{equation}
		\chi\left(\,\spindle\,\right) = 2-\left(1-\f{1}{n_-}\right)-\left(1-\f{1}{n_+}\right) = \f{n_-+n_+}{n_-n_+}\,,
	\end{equation}
	while the total R-symmetry flux through the spindle is given by
	\begin{equation}
		\f{1}{2\pi}\int_{\sspindle} F_R = -\chi\left(\,\spindle\,\right)\,.
	\end{equation}
	where $F_R = \dd A^{(1)}+\dd A^{(2)}$. Hence, this solution realises the twist scenario of \cite{Ferrero:2021etw}. Let us re-emphasise that this is not a standard topological twist. The spindle is a so-called "bad" orbifold which does not allow for a constant curvature metric as one would expect from a topological twist. Similarly, the Killing spinor is not constant along the spindle. This solution only topologically realises the topological twist. As already mentioned, in order to find solutions with global spindle topology we need $q_1q_2<0$ analysing the fluxes one can see that this implies that also $p_1 p_2 <0$. Since from the twist condition we have $p_1 + p_2 = n_+ + n_-$ we have to impose either $p_1<0$ or $p_1> n_+ + n_-$. 
	
	So far we described the local form of the solution as well as the global completion of the two-dimensional surface to a spherical topology. In the case where $\spindle$ is a sphere this is a complete description and with the data provided all $\UU(1)$ bundles over $\spindle$ are uniquely determined. However, as explained in \cite{Ferrero:2021etw}, on orbifolds there is a certain ambiguity left for the $\UU(1)$ orbibundles. More precisely, in the neighbourhood of the conical defects our solution is a $\UU(1)^2$ orbibundle over $\bC/\bZ_n$. On the covering space $\bC\times S^1 \times S^1$ we can introduce complex coordinates $(x,y_i)$ with $|y_i|=1$ such that the generator of $\bZ_n$ acts on $x$ as $x\mapsto \omega x$, with $\omega= \exp\left(\f{2\pi \ii}{n}\right)$. To fully specify the orbibundles over $\bC/\bZ_n$, we need to pick a homomorphism $\varphi$ from the orbifold group into the fibre group,
	\begin{equation}
		\varphi \quad:\quad \bZ_n \rightarrow \UU(1)^2 \quad: \quad \omega \mapsto \varphi(\omega) = \left( \omega^{m^{(1)}},\omega^{m^{(2)}}\right)\,.
	\end{equation}
	This choice of homomorphism amounts to picking `charges' $m^{(i)}\in \bZ_n$ for the $\bZ_n$ action on the fibre. The orbibundle is then fully determined as the quotient $(\bC\times \UU(1)^2)/\bZ_n$ where $\bZ_n$ acts with generator $(\omega,\omega^{m^{(1)}},\omega^{m^{(2)}})$. Hence, we need to supply the solution above with 2 pairs of integers $(m^{(1)}_{\pm},m^{(2)}_{\pm})$ specifying the homomorphism at the two orbifold singularities. In a patch around the singular points we can then write the gauge fields as 
	\begin{equation}\label{eq:conicalgauge}
		\left. A^{(i)}\right|_{w\rightarrow w_\pm} = A^{(i)}_{\pm,0} + \f{m^{(i)}_{\pm}}{n_\pm} \dd z\,,
	\end{equation}
	where $A^{(i)}_{\pm,0}$ is a well-defined gauge field in the disc centred at the conical defect, i.e. $A^{(i)}_{\pm,0}(w_\pm) = 0$. This choice fixes the remaining gauge freedom in \eqref{eq:7dM5sol} and will be crucial for a proper analysis of the singularities in eleven dimensions. Finally, in order to ensure supersymmetry, we need to carefully analyse the regularity of the spinor at the conical defects. As Shown in \cite{Ferrero:2021etw}, in the gauge used in \eqref{eq:7dM5sol} the Killing spinor can be written as
	\begin{equation}
		\zeta = \begin{pmatrix}
			\zeta_+ \\ \zeta_-
		\end{pmatrix} = \e^{-\f{3\ii z}{4}}\f{y^{\f1{20}}}{\sqrt 2H^{\f15}}\begin{pmatrix}
			\left[ \sqrt H + 2 y^{\f32} \right]^{\f12} \\
			\left[ \sqrt H - 2 y^{\f32} \right]^{\f12}
		\end{pmatrix}
	\end{equation}
	From \eqref{eq:FHfuncs} we immediately see that $\zeta_-$ vanishes at both conical defects, hence the focus will be on $\zeta_+$. To properly analyse the regularity at the conical defect it is instructive to change gauge to \eqref{eq:conicalgauge} and furthermore perform a large gauge $\SL(2,\bZ)$ gauge transformation removing the fractional term. Since the spinor has charge $q = \f12$ under the R-symmetry we can rewrite $\zeta_+$ near a conical defect as
	\begin{equation}
		\zeta_+ = \e^{\ii\left(q - \f{1+m_\pm^{(1)}+m_\pm^{(2)}}{2n_\pm} \right)z} f_\pm(r)\,.
	\end{equation}
	where $f_\pm(r)$ is a function of the radial coordinate only. Since at the conical defect the coordinate $z$ is not well-defined, we need to impose
	\begin{equation}
		q - \f{1+m_\pm^{(1)}+m_\pm^{(2)}}{2n_\pm} = 0 \qquad \Leftrightarrow \qquad 1+m_\pm^{(1)}+m_\pm^{(2)} = n_\pm\,,
	\end{equation}
	in order for the spinor to be well-defined and the solution to preserve supersymmetry globally. As we will see below, in the uplift this condition precisely translates into the condition for the internal manifold to be Calabi-Yau, as expected for a supersymmetric solution.

	\section{Uplift and analysis}                                %
	\label{sec:uplift}                                           %
	
	A fruitful strategy to better understand the properties of these solutions, as well as their holographic dual SCFTs, is to consider the uplift to eleven-dimensional supergravity. The general formulae to uplift a solution of maximal 7d gauged supergravity were derived in \cite{Nastase:1999kf,Nastase:1999cb,Cvetic:1999xp} and can be found in Appendix \ref{app:upliftformulae}. 
	
	To attain a more elegant presentation it is convenient to introduce the following functions
	\begin{equation}
		\begin{aligned}
			\Upsilon_H = \f{\mu_0^2 \,H}{w} + w\left( \mu_2^2\, h_1 + \mu_1^2\, h_2 \right)\,,\\
			\Upsilon_F =  \f{\mu_0^2\,f}{4w^3} +  \f{1}{4w}\left( \mu_2^2 \,h_1 + \mu_1^2\, h_2 \right)\,,\\
			\Psi = \mu_0^2\f{\mu_2^2-\mu_1^2}{\mu_2^2+\mu_1^2} + \f{w^2}{H} \left( \mu_2^2\, h_1 - \mu_1^2\, h_2 \right)\,,\\
		\end{aligned}
	\end{equation}
	where we introduced the embedding coordinates $\mu_I$ on the four-sphere which satisfy $\mu_0^2+\mu_1^2+\mu_2^2 = 1$. A convenient choice for our purposes is, 
	\begin{equation}
		\mu_0 = \cos\xi\,,\qquad \mu_1 = \sin\xi\cos\theta\,,\qquad  \mu_2 = \sin\xi\sin\theta\,.
	\end{equation}
	where $\xi\in[0,\pi]$ and $\theta\in [0,\pi/2]$. In terms of these function we can write the eleven-dimensional metric as
	\begin{equation}
		\begin{aligned}
			L^{-2} \ds_{11}^2 = \Upsilon_H^{1/3} \Bigg[ \ds_{\AdS_5}^2 + \ds_{\sspindle}^2 + \f{1}{\Upsilon_H}\Big( w^2 \dd\mu_0^2 &+ h_1 \,\dd\mu_1^2 + h_2\,\dd\mu_2^2 \\
			&+ \mu_1^2\, h_1 \,\DD\phi_1^2 + \mu_2^2\, h_2 \,\DD\phi_2^2\Big) \Bigg]
		\end{aligned}
	\end{equation}
	where we introduced the gauged one-forms
	\begin{equation}\label{eq:phiforms}
		\DD\phi_i = \dd\phi_i - A^{(i)}\,,
	\end{equation}
	and the coordinates $\phi_i$ are both $2\pi$-periodic. This metric is supported by the four-form flux $G_4 = \dd C_3$ with three-form potential, 
	\begin{align}
		L^{-3} C_3 =& -\f{\Psi}{2w\,\Upsilon_H}\DD\phi_1\wedge \DD\phi_2\wedge\dd\mu_0 - \f{\mu_0\mu_1^2\mu_2^2}{(\mu_1^2+\mu_2^2)}\f{1}{w\,\Upsilon_H}\DD\phi_1\wedge \DD\phi_2\wedge \dd\left(\log\f{\mu_1}{\mu_2}\right)\nn\\
		&- \mu_0\, \sqrt{w H} \left( \f{q_1}{h_1^2}\dd\phi_2 + \f{q_2}{h_2^2}\dd\phi_1 \right)\wedge \dvol_{\sspindle}\,.\nn
	\end{align}
	These solutions are a family of $\f14$-BPS solutions of eleven-dimensional supergravity and therefore should fall into the general classification of \cite{Gauntlett:2004zh}. Although conceptually straightforward, finding an explicit map to the canonical GMSW form is very tedious. It remains a fruitful task however since the many of the computations simplify in the classification form. The transformation to canonical coordinates is relegated to Appendix \ref{app:GSMW} to shield the reader from some eyesore. 
	
	\subsection{Regularity in eleven dimensions}    %
	
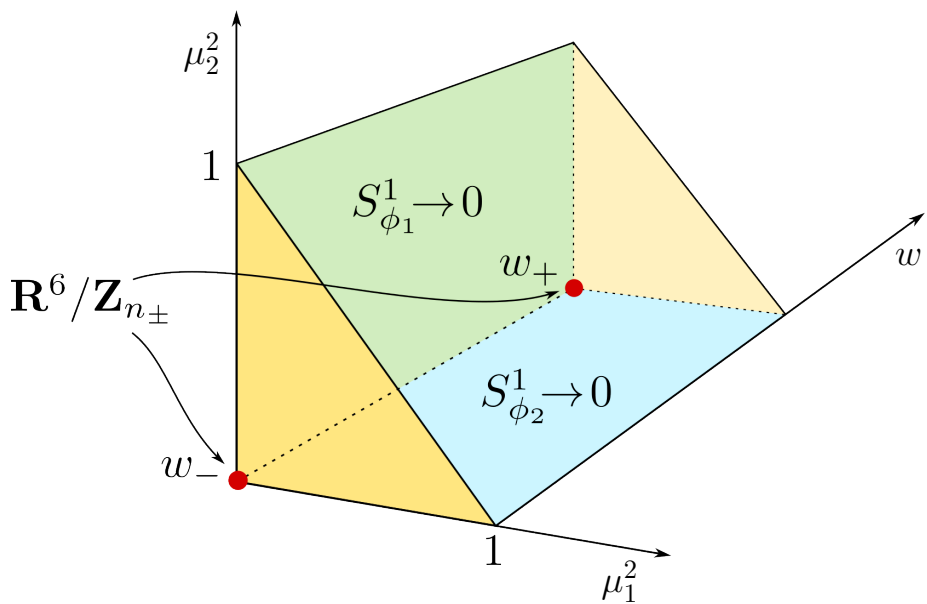
\begin{figure}[!htb]
		\centering
		\begin{tikzpicture}
			\node (myPic) at (0,0) {\includegraphics[width=0.75\textwidth]{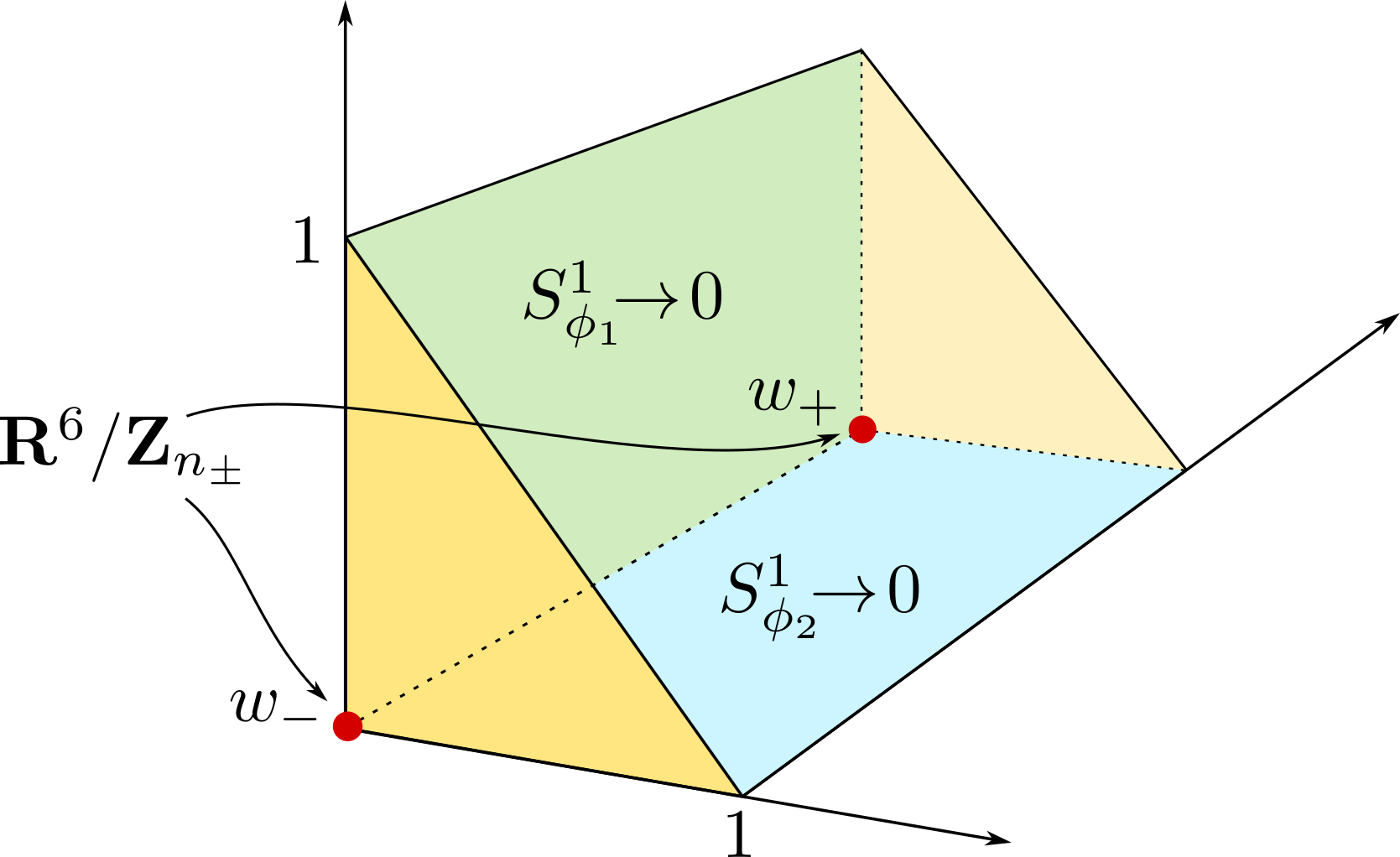}
			};
			\node at (2,-3.9) {\Large$\mu_1^2$};
			\node at (-3.5,3.1) {\Large$\mu_2^2$};
			\node at (5.8,0.4) {\Large$w$};
		\end{tikzpicture}
		\caption{A schematic representation of the prism over which the three circles are fibred. In the interior of the prism, all fibres remain of finite size. However, along the faces a variety of cycles shrink as indicated on the figure. The red dots in the corners indicate the presence of $\bR^6/\bZ_{n_\pm}$ singularities.}
		\label{fig:prism}
	\end{figure}

	As discussed above, the seven-dimensional solution contains two conical singularities, one at each pole of the spindle. After uplifting the solution, let us investigate the fate of these singularities in eleven dimensions. The eleven-dimensional metric takes the form
	\begin{equation}
		\ds_{11}^2  = \e^{2\lambda}\left( \ds_{\AdS_5}^2 + \ds_{\cM_6}^2\right)\,,  
	\end{equation}
	where we can regard the six-dimensional internal space $\cM_6$ as a U$(1)^3$ fibration over a prism with coordinates $(w,\mu_1, \mu_2)$, represented graphically in Figure \ref{fig:prism}.

	Along the diagonal face where $\mu_0=0$, or equivalently $\mu_1^2 + \mu^2_2=1$, and away from the faces boundary no circle shrinks. Instead, along this face for generic $w$, the internal space looks like $\mathbb{R}\times S^3\ltimes \spindle$ where the $S^3$ is located at the equator of the $S^4$. This is nothing but the usual degeneration of an $S^4$, written as an $S^3$-fibration over an interval. The $S^3$ shrinks at $\mu_{1}^2=\mu_2^2=0$, or equivalently at $\mu_0=\pm 1$. Note that this diagram only represents half of the physical internal space, since we have $\mu_0\in[-1,1]$ and the diagram only represents the positive values of $\mu_0$. The negative values can be obtained by reflecting the diagram along the diagonal face. Note that this means that also the singularities are doubled. Along the face where $\mu_i=0$, the $S^1_{\phi_i}$ circle shrinks, these are just the poles of the $S^3$. Both circles shrink smoothly, provided they are $2\pi$-periodic. 
	
	The most delicate points to analyse are the simultaneous limits $\mu_1,\mu_2\rightarrow 0$ and $w \rightarrow w_\pm$, in figure \ref{fig:prism} these are the red dots.\footnote{The analysis for $\mu_0=\pm 1$ is identical, so below we do not specify which one we pick.} To properly do so we should be careful with our choice of gauge and in particular keep track of large gauge transformations. The appropriate choice for the gauge fields around the singular points is given by \eqref{eq:conicalgauge} so for the analysis in this section we substitute this expression in \eqref{eq:phiforms}. To continue our analysis in the neighbourhood of the singular points, let us redefine
	\begin{equation}
		\sin \xi=\frac{2\, r \, \cC\, w_\pm \cos\chi}{\sqrt{\cos^2\theta \,h_1(w_\pm)+\sin^2\theta \,h_2(w_\pm)}}\, ,\quad  w=w_\pm + r^2\frac{\cC^2 \, f'(w_\pm) \,\sin^2\chi}{4 w_\pm}\, ,
	\end{equation}
	where $\cC$ is an arbitrary positive constant. Notice we have $f'(w_+)<0$ and $f'(w_-)>0$ so the change of coordinates is well defined at both endpoints. 
	Taking the $r\rightarrow 0$ limit (and performing the correct large gauge transformations) we find that the metric takes the form:
	\begin{equation}
		\dd s^2_{\mathcal{M}_6}=\dd \mu_1^2+\mu_1^2\left(\dd \phi_1+\frac{m^{(1)}_{\pm}}{n_\pm}\dd z\right)^2+\dd \mu_2^2+\mu_2^2\left(\dd \phi_2+\frac{m^{(2)}_{\pm}}{n_\pm}\dd z\right)^2+\dd \mu_3^2+\frac{\mu_3^2}{n_{\pm}^2}\dd z^2\, ,
	\end{equation}
	where 
	\begin{equation}
		\begin{split}
			\mu_1&=\frac{r \cos\chi\sqrt{h_1(w_{\pm})}\cos\theta}{\sqrt{h_1(w_{\pm})\cos^2\theta+h_2(w_{\pm})\sin^2\theta}}\, ,\quad \mu_2=\frac{r \cos\chi\sqrt{h_2(w_{\pm})}\sin\theta}{\sqrt{h_1(w_{\pm})\cos^2\theta+h_2(w_{\pm})\sin^2\theta}}\, ,\\
			\mu_3&=r \sin\chi\, .
		\end{split}
	\end{equation}
	Staring at this expression for a while, one notices that this is nothing but $\bR^6/\bZ_{n_{\pm}}$. In other words, we find that the $\bR^2/\bZ_{n_\pm}$ orbifold singularities of the seven-dimensional solution are lifted to $\bC^3/\bZ_{n_\pm}$ singularities in eleven dimensions at the poles of the $S^4$, \cite{BenettiGenolini:2023ndb}. More precisely, the action on the complex coordinates, $(z_i=\mu_i\e^{\ii \psi_i})$ is given as follows,
	\begin{equation}\label{eq:orbaction}
		\bigg( z_1\, , \,z_2\, , \,z_3 \bigg) \rightarrow \left( \e^{2\pi\ii \f{m^{(1)}_\pm}{n_{\pm}}} z_1\, , \, \e^{2\pi\ii\f{m^{(2)}_\pm}{n_{\pm}}} z_2\, , \,\e^{2\pi\ii\f{1}{n_{\pm}}}z_3 \right)\,.
	\end{equation}
	Later we will need to know the weights of the orbifold action with respect to the R-symmetry vector. From the orbifold action \eqref{eq:orbaction}, it is simple to read off the weights to be:
	\begin{equation}\label{eq:orbweights}
		\epsilon^{(i)}_{\pm}=\frac{1}{n_{\pm}}\left(m_{\pm}^{(1)},m_{\pm}^{(2)},1\right)\, .
	\end{equation}
	An alternative method to see that these are indeed the weights is to compute the anti-symmetric matrix: $\tfrac{1}{2}(\dd\xi^{\#})_{\mu\nu}$ where $\xi^{\#}$ is the one-form dual to the R-symmetry vector $\xi$. By writing this in canonical anti-symmetric matrix form one can read off the weights as the off-diagonal entries. 
	Computing this matrix at the fixed points we find:
	\begin{equation}
		\frac{1}{2}\dd\xi^{\#}_{\mu\nu}=\frac{1}{n_{\pm}}\begin{pmatrix}
			0&m_{\pm}^{(1)}&0&0&0&0\\
			-m_{\pm}^{(1)}&0&0&0&0&0\\
			0&0&0&m_{\pm}^{(2)}&0&0\\
			0&0&-m_{\pm}^{(2)}&0&0&0\\
			0&0&0&0&0&1\\
			0&0&0&0&-1&0
		\end{pmatrix}_{\mu\nu}\, ,
	\end{equation}
	which indeed gives the same weights as in \eqref{eq:orbweights}.

	Summarising, we find that locally the eleven-dimensional metric takes the form
	\begin{equation}
		\ds_{11}^2 = \ds^2_{\AdS_5} + \ds_{\bC^3/\bZ_{n_\pm}}^2\,.
	\end{equation}
	with the orbifold action given in \eqref{eq:orbaction}. This is in contrast to the $\cN=2$ preserving class $\cS$ punctures that locally are of the form $\AdS_5 \times S^2 \times \bC^2/\bZ_n$. Hence, we find that the conical defects of the spindle solution realise genuine locally $\cN=1$ preserving punctures! Upon setting one of the parameters $m^{(i)}$ to zero, we recover the more familiar locally $\cN=2$ preserving punctures. In this case one complex plane remains untouched by the orbifold action and the corresponding $\UU(1)$ is enhanced to the $\SU(2)$ R-symmetry. Indeed, we see that in this case the two-dimensional orbifold acts as $\f1n(-1,1)$, and our punctures reduce to the well-known type $A$ du Val singularities. I.e. in this case the local singularities are classified by (collections of) finite subgroups of $\SU(2)$ while in the $\cN=1$ case they are classified by finite subgroups of $\SU(3)$. The latter classification \cite{Fairbairn:1964sga} is much richer and our spindle solutions only realise a small subset of them given by the abelian orbifolds of the form $\bC^3/\bZ_{n_\pm}$. It is an outstanding problem to find more general solutions realising punctures corresponding to the remaining finite subgroups of $\SU(3)$. We hope to come back to this problem in the near future.

	\subsection{Flux Quantisation}    %
	
	In order for the backgrounds described above to be consistent backgrounds of M-theory, we impose that the $G_4$ flux is quantised along any four-cycle in $\cM_4$. 
	
	The first non-trivial cycle we consider, which we denote by $\cC_4$, has the topology of $S^4$ and can be obtained as the $S^4$ fibre over a generic point on the spindle. This four-cycle can be obtained by taking a slice of the prism perpendicular to the $w$-axis in Figure \ref{fig:prism} and integrating along the both $\dd\phi_i$ circles. Using the expression for the four-form flux above, we have the following quantisation condition
	\begin{equation}
		\f{1}{(2\pi\ell_p)^3}\int_{\cC_4} G_4 = N \in \bZ
	\end{equation}
	This flux can be interpreted as the number of M5-branes wrapping the spindle. As a consequence, we find the length scale $L$ of the uplifted solution is quantised as
	\begin{equation}\label{eq:Nquant}
		L = (\pi N)^{\f13}\ell_p\,.
	\end{equation}
	Next, consider the four-cycles $\cD_4^{\pm}$ obtained by going to one of the orbifold points at the poles of the spindle. The fibres over these points take the form $S^4/\bZ_{n_\pm}$. Similar to the above we can consider the flux quantisation through the fibres, 
	\begin{equation}
		\f{1}{(2\pi\ell_p)^3}\int_{\cD_4^{\pm}} G_4 = \f{N}{n_\pm}\,.
	\end{equation}
	The cycles $\cD_4^{\pm}$ are not independent from the cycle $\cC_4$ but instead we have the homology relations
	\begin{equation}
		\left[\cC_4\right] = n_\pm \left[\cD_4^{\pm}\right]\,.
	\end{equation}
	This agrees with the analysis of \cite{BenettiGenolini:2023ndb} using equivariant localisation.
	
	\section{Wrapped M2- and M5-branes}      %
	\label{sec:observables}                  %
	
	Probe M5- and M2-branes wrapping cycles of the internal manifold give rise to an important set of protected data in the dual field theory. Probe BPS M2-branes wrapping calibrated two-cycles correspond to chiral primaries in the dual SCFTs \cite{Gauntlett:2006ai}, while the probe BPS M5-branes can be understood as parametrising a class of supersymmetric deformations of the four-dimensional $\cN=1$ SCFT. In this section we study the possible embeddings of BPS probe branes within the spindle solution. We will find that there is a richer structure of probe branes that may be embedded in comparison with the BBBW solutions \cite{Bah:2011vv} as studied in \cite{Bah:2013wda}. In particular, on top of the BPS M5-branes which deform the theory by changing/adding a puncture, as in \cite{Bah:2013wda}, there are additional M5 branes wrapping the isometry circle of the spindle. These are located away from the conical defects and provide a new set of codimension-1 operators in the 6d $(2,0)$ which deform the resulting four-dimensional theory. One can attribute the presence of these additional probe M5-branes to a larger moduli space for the probe M5-branes in the setup and therefore a richer structure in the dual SCFT. 
	
	The analysis of \cite{Bah:2013wda} was performed by embedding their setup into the GMSW framework, and exemplified with the BBBW \cite{Bah:2011vv} and $\cN=1$ Maldacena--Nunez solutions \cite{Maldacena:2000mw}. After embedding our solution into this general framework, which is detailed in Appendix \ref{app:GSMW}, the analysis can be repeated verbatim. However, due to the richer setup our results include a larger set of (new) probe branes.
	
	\subsection{Probe M5 branes}
	
	In order for a probe M5-brane to preserve the symmetries of our solutions it is forced to wrap all of AdS$_5$ together with a one-cycle in the internal space. Let $\{\sigma^{i}\}$ with $i=0,\dots, 4$ be the coordinates in AdS$_5$ and let $\tau=\sigma^5$ be the proper length parameter along the one-cycle within $M_6$ wrapped by the probe M5-brane. The calibration condition for a probe M5-brane fully wrapping AdS$_5$, written using the canonical $\mathcal{N}=1$ form (see appendix \ref{app:GSMW}), is
	\begin{equation}\label{eq:calibration}
		\e^{-3\lambda}\dd s_4\Big|_{\dd\tau}=0\, ,\qquad \frac{\e^{-3\lambda}}{\cos\zeta}\dd y\Big|_{\dd\tau}=0\, ,\qquad \frac{\sin\zeta \cos\zeta}{3}(\dd \psi+P)\Big|_{\dd\tau}=0\, ,
	\end{equation}
	where $\dd s^2_4$ is the ($y$-dependent) K\"ahler metric at a fixed value of $y$. The component $g_{\tau\tau}$ of the induced metric is
	\begin{equation}
		g_{\tau\tau}=\e^{-6\lambda}\left[\dd s_4\Big|_{\dd\tau}\right]^2+\frac{\e^{-6\lambda}}{\cos^2\zeta}\left[\dd y|_{\dd\tau}\right]^2+\frac{\cos^2\zeta}{9}\left[(\dd \psi+P)\Big|_{\dd\tau}\right]^2\, ,
	\end{equation}
	which we require to be non-zero. The only option, giving a non-trivial induced metric $g_{\tau\tau}$, and satisfying the calibration condition is to set $\sin\zeta=0$, which is equivalent to $y=0$. This immediately satisfies the second condition in \eqref{eq:calibration} since we are localised to a fixed value of $y$ as well as the third one. It remains to satisfy the first condition in  \eqref{eq:calibration}. From the embedding of the spindle solution in the GMSW classification we have that the coordinate $y$ is given by $y=w\cos\xi$ and therefore we have that $y=0$ if and only if $\cos\xi=0$. This then places us at the equator of the $S^4$ which has the topology of a three-sphere.
	The location of the probe M5-branes is then at a point on the diagonal face of figure \ref{fig:prism}. For ease of reference let us denote this face by $\cF_2$. 
	
	As the metric on $M_6$ is toric, it admits a $\UU(1)^3$ action which is fibered over the prism in figure \ref{fig:prism}. The M5-brane will wrap a one-cycle embedded in this torus and we must require that the one-cycle closes, i.e. that it is not dense within the torus. At generic points on the face $\cF_2$ the M5-brane will then break a $\UU(1)^2\subset \UU(1)^3$. The preserved symmetry is enhanced along the boundary of $\cF_2$ where an additional cycle shrinks and therefore this symmetry is not broken by the presence of the probe brane.
	
	Once we have restricted to the equator of the four-sphere, it remains to satisfy the first of calibration condition, \eqref{eq:calibration}. It is most convenient to work in terms of a vielbein on $\dd s^2_4$, which we derive in appendix \ref{app:GSMW} in terms of the canonical GMSW form. These vielbein are adapted to give the K\"ahler form on the base in the canonical way. The pertinent vielbein are the ones involving the angular coordinates on the four-dimensional base, which evaluated at $y=0$ and along the curve, are given by
	\begin{equation}
		\begin{split}
			e^2\Big|_{y=0}&= \frac{\cos\theta\sin\theta }{\sqrt{ \Theta(w,\theta)}}\left(h_1(w)\frac{\dd\phi_1}{\dd \tau}-h_2(w)\frac{\dd\phi_2}{\dd\tau}\right)\dd \tau\, ,\\
			e^4\Big|_{y=0}&=-\frac{ \sqrt{f(w)}}{2\sqrt{\Theta(w,\theta)}}\left(\cos^2\theta\frac{\dd\phi_1}{\dd\tau}+\sin^2\theta\frac{\dd\phi_2}{\dd\tau}\right)\dd\tau\, .
		\end{split}
	\end{equation}
	To solve \eqref{eq:calibration} these must simultaneously vanish. 
	
	\subsubsection*{Generic point in \texorpdfstring{$\cF_2$}{F2}}
	
	At a generic point $(w_*,\theta_*)\in \cF_2$ we must impose the two conditions
	\begin{equation}
		\frac{\dd\phi_1}{\dd\tau}=\frac{h_2(w_*)}{h_1(w_*)}\frac{\dd\phi_2}{\dd\tau}\, ,\qquad   \frac{\dd\phi_1}{\dd\tau}=-\tan^2\theta_* \frac{\dd\phi_2}{\dd\tau}  \, .
	\end{equation}
	Since $h_i(w_*)>0$ the only possibility is 
	\begin{equation}
		\frac{\dd\phi_1}{\dd\tau}= \frac{\dd\phi_2}{\dd\tau}=0 \, .
	\end{equation}
	Hence, the probe M5-brane is located at a generic point along the $\phi_i$ circles. That is, at a generic point on the $S^3$ along the equator of the $S^4$. The cycle wrapped by the probe brane is then simply the isometry circle of the spindle, 
	\begin{equation}
		\text{M5}_{\text{gen}}: ~~\text{AdS}_5\times S^{1}_{\text{spindle}}\times \{\text{point}\in S^3_{\text{equator}}\}
	\end{equation}
	Since the brane is localised at points along the $\phi_i$ circles it follows that the $\UU(1)^3$ symmetry is broken to $\UU(1)_R$ by the probe M5-brane, with the remaining isometry dual to the R-symmetry. 
	
	\subsubsection*{North and south pole of \texorpdfstring{$S^3_{\text{equator}}$}{S3equator}}
	
	A more interesting case arises when the probe brane approaches the boundaries of this domain, i.e. when it is located at a pole of the three-sphere. Consider first the north pole of the $S^3_{\text{equator}}$ at $\mu_1=0$, where the circle with coordinate $\phi_1$ shrinks. Note that we stay away from $w=w_{\pm}$. In this case, the calibration conditions reduces to
	\begin{equation}
		\frac{\dd \phi_2}{\dd\tau}=0\, ,
	\end{equation}
	such that the probe M5-brane is localised at a point along the $\phi_2$ circle. This is just the non-shrinking circle $S^1_N$ of the north pole of $S^3_{\text{equator}}$. We see that compared to the generic case there is an enhancement of the symmetry preserved by the probe brane to $\UU(1)^2$, with the two $\UU(1)$'s being the R-symmetry and a flavour symmetry. The boundary at $\mu_2=0$ works similarly, this time with the M5-brane localised at point along the $\phi_1$ circle, that is $S^1_{S}$. We denote these M5-brane probes by
	\begin{equation}\label{eq:M5NS}
		\text{M}5_{w}^{N,S}=\text{AdS}_5\times S^1_{\text{spindle}}\times \{w\in(w_-,w_+)\}\times \{S^3_{N,S},~\text{point}\in S^1_{N,S}\}\, .
	\end{equation}
	Both the $\text{M5}_{\text{gen}}$ and $\text{M}5_{w}^{N,S}$ probe M5-branes wrap the circle of the spindle and are located at points of the equator of the $S^4$. These probe M5-branes are new in comparison with the possible probe M5-branes possible in the compactification on a Riemann surface and is an artefact of the need to twist the R-symmetry with the isometry of the spindle. Finally observe that there is no obstruction to moving a $\text{M5}_{\text{gen}}$ to $\text{M}5_{w}^{N,S}$ or vice-versa.

	\subsubsection*{Spindle fixed points}
	
	The remaining possibility is to put the probe M5-branes at the poles of the spindle $w=w_\pm$. These are even more interesting and we will treat them concurrently. At these points the $\UU(1)$ isometry circle of the spindle shrinks and is therefore not wrapped by the probe M5-branes. Away from the poles of the $S^3_{\text{equator}}$ the embedding of the probe brane must satisfy
	\begin{equation}
		\frac{\dd\phi_1}{\dd\tau}=\frac{h_2(w_\pm)}{h_1(w_\pm)}\frac{\dd\phi_2}{\dd\tau}\, .
	\end{equation}
	In order for the one-cycle not to be dense, which would render it inappropriate for the probe M5-brane to wrap, we must require that 
	\begin{equation}\label{eq:rationalcondition}
		\frac{h_2(w_\pm)}{h_1(w_\pm)}
		\equiv\frac{\alpha_{2,\pm}}{\alpha_{1,\pm}}\in\bQ\, ,\qquad \text{where}\quad \alpha_{i,\pm}\in \bZ\, .
	\end{equation}
	At the poles $w=w_\pm$ we have the embedding,
	\begin{equation}\label{eq:M5pm}
		\phi_1(\tau)=\phi_{1,*}+\alpha_{2,\pm}\tau\, ,\quad  \phi_2(\tau)=\phi_{2,*}+\alpha_{1,\pm}\tau\, ,
	\end{equation}
	where $\phi_{i,*}$ are the initial point of the one-cycle inside $S^3_{\text{equator}}$. These M5-branes therefore wrap
	\begin{equation}
		\text{M}5_{\pm}^{\vec{\alpha}_{\pm}}: \AdS_5\times \left\{ S^1_{R}(\vec{\alpha}_{\pm})\subset S^3_{\text{equator}}\right\}\times \left\{ w=w_\pm \right\} \, .
	\end{equation}

	For generic solutions to the calibration condition, the parameters $\alpha_{i,\pm}$ in \eqref{eq:rationalcondition} are not integer and additional constraints on the spindle parameters need to be imposed. Following \cite{Ferrero:2021wvk} we may solve for the roots and the parameters $q_i$ in terms of the integer magnetic charges and orbifold weights. We have\footnote{Note that our naming of the orbifold orders is interchanged with that in \cite{Ferrero:2021wvk}.}
	\begin{equation}
		\begin{split}
			w_{\pm}&=\frac{3 p_1 p_2(4 n_{\mp}-n_{\pm}+s)(s+p_1+p_2)}{2(n_{\pm}-p_1)(n_{\pm}-p_2)[s+2(p_1+p_2)]^2}\, ,\\
			q_1&=\frac{3 p_1 p_2^2(5 n_--n_++s)(5 n_+-n_-+s)(p_1-2p_2-s)(p_1+p_2+s)^2}{4(n_--p_1)^2(n_--p_2)^2[s+2(p_1+p_2)]^4}\, ,\\
			q_2&=q_1\Big|_{p_1\leftrightarrow p_2}\, ,
		\end{split}
	\end{equation}
	where
	\begin{equation}\label{eq:sparam}
		s=\sqrt{(n_-+n_+)^2 +12 (n_--p_1)(n_+-p_1)}\, .
	\end{equation}
	For regularity one should take  $n_+>n_->0$, fix $p_2=n_++n_--p_1$ and take either $p_1<0$ or $p_1>n_++n_-$. It then follows that \eqref{eq:rationalcondition} is true if and only if the parameter $s$ in equation \eqref{eq:sparam} is rational. This also implies that such SCFTs have a rational central charge (at leading order). Note that a similar phenomenon was found for the BBBW solutions in \cite{Bah:2013wda} where the rational SCFTs allowed additional probe M5-branes at a point on the Riemann surface, wrapping a non-trivial one-cycle on the $S^3_{\text{equator}}$. The rationality condition is equivalent to
	\begin{equation}
		12 n_+n_-=m^2 -(n_+-n_-)^2+12 p_1(n_++n_--p_1)\, ,
	\end{equation}
	with $m$ an arbitrary integer. We may solve this by defining
	\begin{equation}
		n_s=n_++n_-\, ,\quad n_{p}=n_+n_-\, ,
	\end{equation}
	such that
	\begin{equation}
		n_p=(p_1-n_s)p_1+ \frac{m^2-n_s^2}{12}\, .
	\end{equation}
	In this way we see that we require $(m,n_s)$ to satisfy the generalised Pell's equation:
	\begin{equation}\label{eq:pell}
		m^2- n_s^2=12 \delta \, ,\qquad \delta \in \bZ\, .
	\end{equation}
	To simplify let 
	\begin{equation}
		n_s=12 k_1+a\, ,\qquad m=12 k_2+b\, ,\qquad\text{where}\qquad k_{1},k_{2}\in\bZ,\qquad a,b\in\bZ_{12}\, ,
	\end{equation}
	then \eqref{eq:pell} can be solved by picking $a$ and $b$ from the sets
	\begin{equation}
		\begin{split}
			(a,b)\in\{0,6\}\, ,~~
			(a,b)\in\{1,5,7,11\}\, ,~~
			(a,b)\in\{2,4,8,10\}\, ,~~
			(a,b)\in\{3,9\}\, .
		\end{split}
	\end{equation}
	One now needs to invert to solve for $n_{\pm}$. We have not found a closed form expression for this in general however we have checked that solutions can be found. Mathematica's function `FindInstance' almost always finds a solution for random integer values consistent with the regularity that we have checked. 
	
	For rational spindle SCFTs, we may wrap an additional probe M5-brane along a non-trivial cycle within $S^3_{\text{equator}}$, localised at the poles of the spindle. The wrapped one-cycle depends on a pair of integers $\alpha_{i,\pm}$ which exist for all rational SCFTs. The symmetry preserved in this case is $\UU(1)^2$, with one $\UU(1)$ the R-symmetry and the other a flavour symmetry.

	\subsubsection*{Corners of \texorpdfstring{$\cF_2$}{F2}}
	
	The final set of points on $\cF_2$ that we need to discuss are the corners. For concreteness, let us consider the corner corresponding to the north pole of the $S^3_{\text{equator}}$ and the $w=w_+$ pole of the spindle. At this point the calibration condition is immediately solved and the one-cycle wrapped is given by the $S^{1}_{N}$ of the north pole. The generalisation to the other corners is obvious hence these probe M5-branes wrap
	\begin{equation}\label{eq:F2cornersM5}
		\text{M5}_{\pm}^{N,S}:\text{AdS}_5\times S^{1}_{N,S}\times (\spindle^{\pm})\times(S^3_{\text{equator}})^{N,S}\, .
	\end{equation}
	These M5-branes preserve the full $\UU(1)^3$ symmetry. 
	
	Consider now moving one of the $\text{M5}^{\pm}_{\alpha_\pm}$ towards a neighbouring corner. For concreteness let us take the $w=w_+$ pole of the spindle and move the point on the three-sphere to the north pole of $S^3_{\text{equator}}$. Since the circle with coordinate $\phi_1$ shrinks at this point, the M5-branes wraps the one-cycle $S^1_N$ $\alpha_{1,+}$ times. Equivalently, there are $\alpha_{1,+}$ probe M5-branes wrapping $S^1_{N}$. Conversely if we have an $\text{M5}_{\pm}^{N,S}$ probe brane and try to move it away from the corner we cannot unless we have $\alpha_{1,+}$ of them. Doing so the $\alpha_{1,+}$ probe branes at the corner become a single $\text{M5}^{N}_{\vec{\alpha}_+}$ probe brane in the away from the corner. This is a consequence of the corner being a $\bR^4/\bZ_{\alpha_{1,+}}$ orbifold at which we can place so-called fractional M5-branes as we will discuss in more detail later. At the other corners of $\cF_2$ there is an analogous behaviour. This is very similar to the BBBW solutions, \cite{Bah:2013wda}. There the probe M5-branes in our language are, $\text{M5}_{-}^{N,S}$ and $\text{M5}_{-}^{\vec{\alpha}_-}$. We see that the spindle leads to a doubling of these probe M5-branes with different weights $\vec{\alpha}_{\pm}$ at the two poles of the spindle. Finally, note that we cannot move a probe M5-brane from either of the $w=w_{\pm}$ boundaries into the bulk to a different point on the spindle and vice versa. This will lead to a moduli space which is not connected.  
	
	To summarise, we have found that there are two distinct types of probe M5-branes. One class wraps the spindle circle and the other is located at the poles of the spindle and wraps a one-cycle inside the equator of $S^4$. For the M5-branes at the poles of the spindle there are additional probe M5-branes provided that the dual SCFT is rational, which we have shown exist. These additional M5-branes can be moved to the corners of $\cF_2$ and lead to stacks of (fractional) M5-branes localised at the corners of $\cF_2$. 
	
	\subsection{Moduli space of probe M5-branes}
	
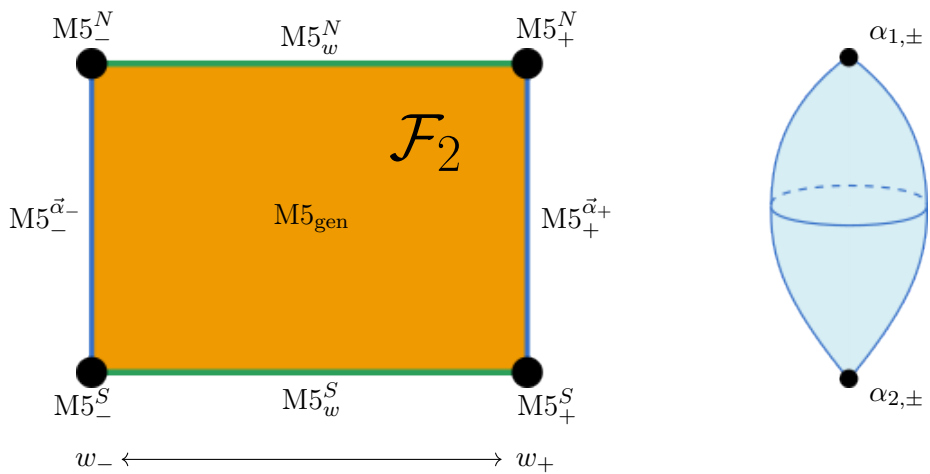
\begin{figure}[!htb]
		\centering
		\begin{tikzpicture}
			\node (myPic) at (0,0) {\includegraphics[width=0.7\textwidth]{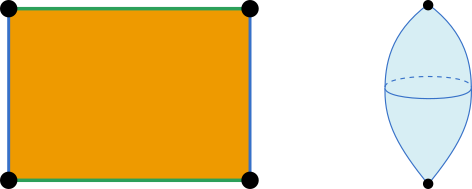}};
			\node at (-1,1) {\Huge $\cF_2$};
			\node at (-2.5,0) {$\text{M5}_{\rm gen}$};
			\node at (-2.5,2.4) {$\text{M5}_w^{N}$};
			\node at (-2.5,-2.4) {$\text{M5}_w^{S}$};
			\node at (-6,0) {$\text{M5}_{-}^{\vec\alpha_-}$};
			\node at (1,0) {$\text{M5}_{+}^{\vec\alpha_+}$};
			\node at (0.6,2.5) {$\text{M5}_{+}^{N}$};
			\node at (-5.5,2.5) {$\text{M5}_{-}^{N}$};
			\node at (0.6,-2.5) {$\text{M5}_{+}^{S}$};
			\node at (-5.5,-2.5) {$\text{M5}_{-}^{S}$};
			\node at (5.2,2.4) {$\alpha_{1,\pm}$};
			\node at (5.2,-2.4) {$\alpha_{2,\pm}$};
			\node at (-5.35,-3.25) {$w_-$};
			\node at (0.45,-3.25) {$w_+$};
			\draw[<->] (-5,-3.2)   -- (0,-3.2); 
		\end{tikzpicture}
		\caption{The moduli space $\cF_2$ of probe M5-branes in out setup. The blue line and four black dots are disconnected from the yellow region and green lines. Moreover, the blue lines are not accessible when the large $N$ central charge of the SCFT is not rational. When the theory is rational the blue line and the two connected dots are topologically a spindle with weights given by $\vec{\alpha}_{\pm}$. }
		\label{fig:moduliM5}    
	\end{figure}

	Having studied the possible embeddings of probe M5-branes we turn our attention to better understanding the moduli space of the probe M5-branes. We mainly focus on the cases satisfying the rationality condition, since these present the richest set of probe M5-branes. When this condition is not satisfied part of the moduli space, coloured in blue in Figure \ref{fig:moduliM5}, does not exist.  
	
	Since the probe M5-branes are calibrated by the R-symmetry direction at $y=0$ the metric on the moduli space is obtained by restricting the metric on $\mathcal{M}_4$ to $\cF_2$. We take the $\alpha_{i,\pm}$ as defined in \eqref{eq:rationalcondition} and take them to be pairwise relatively prime, i.e. $\gcd(\alpha_{1,\pm},\alpha_{2,\pm})=1$. More explicitly, they are given in terms of the $n_{\pm}$ and $p_i$ by
	\begin{equation}
		\alpha_{1,\pm}=p_2\big(n_-+n_++6(n_\pm-p_2)-s\big)\, ,\quad \alpha_{2,\pm}=p_1\big(n_-+n_++6(n_{\pm}-p_1)-s\big)\, ,
	\end{equation}
	and satisfy
	\begin{equation}
		\alpha_{1+}=6p_2(n_+-n_-)+\alpha_{1-}\, ,\quad \alpha_{2+}=6p_1(n_+-n_-)+\alpha_{2-}\, .
	\end{equation}
	Pulling back the metric on $\mathcal{M}_4$ to the diagonal face $\mathcal{F}_2$ in \ref{fig:prism} we find
	\begin{equation}
		\begin{split}
			\dd s^2_4&=\Theta(w,\theta)\dd\theta^2+\frac{\cos^2\theta\sin^2\theta}{\Theta(w,\theta)}(h_1(w)\dd\phi_1-h_2(w)\dd\phi_2)^2\\
			&+\frac{w^2\Theta(w,\theta)}{f(w)}\dd w^2+\frac{f(w)}{\Theta(w,\theta)}(\cos^2\theta \dd\phi_1+\sin^2\theta\dd\phi_2)^2\,.
		\end{split}
	\end{equation}
	This metric is necessarily complex and K\"ahler. Moreover, it describes a four-dimensional toric orbifold. One should contrast this with the BBBW case where the four-dimensional moduli space is a spindle fibered over a Riemann surface \cite{Bah:2013wda}.
	
	To better understand the moduli space let us construct the degenerating Killing vectors on the boundary of $\cF_2$. For $\mu_1=0,\mu_2=0$ we see that the degenerating Killing vectors, normalised to have unit surface gravity, are respectively
	\begin{equation}
		l_{1}=\partial_{\phi_1}\, ,\quad l_{2}=\partial_{\phi_2}\, .
	\end{equation}
	The metric degenerates smoothly provided the $\phi_i$ have period $2\pi$ which is indeed the period inherited from the four-sphere origin. Since they are $2\pi$-periodic we will use $\partial_{\phi_i}$ as a basis for the $\UU(1)^2$ toric action. In this basis we have the toric vectors:
	\begin{equation}
		v_1=(1,0)\, ,\quad v_3=(0,1)\, .
	\end{equation}
	
	For $w=w_{\pm}$ we see that the degenerating Killing vectors are
	\begin{equation}
		k_{\pm}=\frac{1}{6 n_\mp (n_--n_+)}\left(\alpha_{2,\pm}\partial_{\phi_1}+\alpha_{1,\pm}\partial_{\phi_2}\right)\, ,
	\end{equation}
	where the normalisation is fixed so that the surface gravity is 1. From these we extract out the effective toric vectors:
	\begin{equation}
		v_2=\frac{1}{6 n_-(n_--n_+)}(\alpha_{2,+},\alpha_{1,+})\, ,\quad v_4=-\frac{1}{6 n_+(n_--n_+)}(\alpha_{2,-},\alpha_{1,-})\, .
	\end{equation}
	The non-integral nature of these vectors indicates the orbifolded nature of the degeneration. One can in principle perform a rotation of the basis to obtain an integral set of vectors \cite{Faedo:2024upq} however we will not do this here.\footnote{Note that we have a genuine symplectic structure on our four-dimensional orbifold which is not the case in \cite{Faedo:2024upq}.} 
	
	It follows that at the corners of $\cF_2$ the degeneration is a cover of $\bC^2/\bZ_{\alpha_{i,\pm}}$. For example the metric at the corner $w=w_+$, $\mu_1=0$ corner is:
	\begin{equation}
		\dd s^2_4=\dd r^2+r^2 \left[\dd \xi^2+\frac{\cos^2\xi}{\alpha_{1,+}^2}\left(\alpha_{1,+}\dd\phi_1-\alpha_{2,+}\dd\phi_2\right)^2+\frac{36 n_{-}^2(n_--n_+)^2}{\alpha_{1,+}^2}\sin^2\xi \dd\phi_2^2\right]\, ,
	\end{equation}
	with analogous degeneration at the other corners. The $\mathbb{Z}_{\alpha_{i,\pm}}$, action is non-trivial and generate a finite subgroup of $\UU(2)$ which is not contained in $\SU(2)$. To see this one can compute the weights of the Killing vector $K=\partial_{\phi_1}+\partial_{\phi_2}$ at the fixed point at $r=0$, resulting in
	\begin{equation}
		d K|_{\bC^2/\bZ_{\alpha_{i,\pm}}}=\frac{1}{\alpha_{i,\pm}}\begin{pmatrix}
			0& -6n_{\mp}(n_--n_+)&0&0\\
			6n_{\mp}(n_--n_+)&0&0&0\\
			0&0&0&\alpha_{i+1,\pm}-\alpha_{i,\pm}\\
			0&0&-\alpha_{i+1,\pm}+\alpha_{i,\pm}&0
		\end{pmatrix}
	\end{equation}
	from which the weights can immediately be extracted.
	
	To conclude, we find that the moduli space of the probe M5-branes is a four-dimensional toric K\"ahler orbifold of quadrilateral type. At the corners of the toric polytope the degeneration is $\bC^2/\bZ_{\alpha_{i,\pm}}$. Using the results of appendix \ref{app:BBBW}, and a lot of care, one can take a limit to the moduli space of the BBBW solutions as computed in \cite{Bah:2013wda}.

	\subsection{M2 branes ending on M5 branes}
	
	The presence of the probe M5-branes allows for M2-branes to end on them. Chiral BPS operators in the dual SCFT correspond to M2-branes moving along a geodesic in AdS$_5$ and wrapping a calibrated curve in the internal space. The calibration two-form is \cite{Gauntlett:2006ai}
	\begin{equation}
		Y'=-\sin\zeta \e^{-6\lambda} J + K_1\wedge K_2\, ,
	\end{equation}
	and the calibration condition for the cycle $\cM_2$ becomes
	\begin{equation}
		\dvol(\cM_2)=Y'\Big|_{\cM_2}\, ,
	\end{equation}
	where $\dvol(\cM_2)$ is the volume element on $\cM_2$ pulled back from $\cM_6$. From the form of the metric and calibration condition we see that the calibrated cycles a priori can fall into two classes. They either wrap the $\UU(1)_R$ direction or they do not and are calibrated by $J$. Given a calibrated M2-brane, wrapping the cycle $\mathcal{M}_2$ we may compute the conformal dimension of the dual BPS operator via \cite{Gauntlett:2006ai}
	\begin{equation}
		\Delta(\mathcal{M}_2)=\frac{N}{4\pi}\int_{\mathcal{M}_2}\e^{3\lambda}Y'\, ,
	\end{equation}
	where we have used \eqref{eq:Nquant} and the definition of $Y'$ is given in \eqref{eq:Yprime}.
	
	Consider first the embedding of an M2-brane which does not wrap the R-symmetry direction. Then the calibration condition requires $\sin\zeta=\pm 1$, i.e. we are at restricted to the fixed points of the $\UU(1)_R$ isometry. Since these are genuine fixed points of $\mathcal{M}_6$ and the entire metric shrinks there are no non-shrinking two-cycles which the probe M2-brane can wrap. This differs with the BBBW solutions where one could wrap the Riemann surface and is a consequence of having a genuine $\mathcal{N}=1$ puncture rather than local $\mathcal{N}=2$ punctures.
	
	Having concluded that the probe M2-brane must wrap the $\UU(1)_R$ R-symmetry directions let us consider their embedding in more detail. There are two options. The first consists of the M2-brane world volume wrapping the spindle located at one of the poles of the $S^4$. The second option consists of wrapping a 2-cycle with the topology of a disc, stretching from one of the fixed points to one of the probe M5-branes at $y=0$. 
	
	The first type probe M2-brane was studied in \cite{BenettiGenolini:2023ndb} and we recover their result. Let us parametrise the surface with the coordinates $(\tau,\sigma)$ and adapt $\tau$ so that it parametrises the $\UU(1)_R$ only. Then we have that the brane wraps the line $(w,\xi,\theta)(\sigma)$, with end-points at $\cos\zeta=0$. In order for the calibration condition to hold we need to make the contribution from $J$ to $Y'$ vanish. Clearly evaluating the calibration form at the poles of the $S^4$ satisfies this demand. We denote these cycles by $\Sigma_{N,S}$ and find that the conformal dimension of the dual BPS operators are
	\begin{equation}
		\Delta(\Sigma_{N,S})=\frac{3p_1 p_2}{n_+ n_- (s+2(n_-+n_+))}\, ,
	\end{equation}
	which agrees with equation (3.58) of \cite{BenettiGenolini:2023ndb}. 
	
	Next, we consider the probe M2-branes which end on a probe M5-brane. First we consider the M2-branes which end on a $\text{M5}_{\pm}^{N,S}$, see \eqref{eq:F2cornersM5}. In this case, the BPS condition implies that 
	\begin{equation}
		\sin^4\xi(\sigma)=C\prod_{I=1}^{4}(w(\sigma)-w_I)^{-\tfrac{w_Ih_1(w_I)}{2 f'(w_I)}}
	\end{equation}
	with the $w_I$ being the four roots of $f(w)$ and the constant $C$ is fixed so that at the location of the M5-brane we have $\sin^2\xi=1$. The topology of the M2-brane world volume in the internal space is that of a disc with boundary the M5-brane. Note that we can take the centre of the disc to be at either of the four points where $\cos^2\zeta=0$. The conformal dimension of the corresponding operators when we are fixed to the pole of the spindle is simple to compute. The two-cycle that the M2-brane wraps is then just a linearly embedded two-sphere inside the $S^4$. We denote these cycles by $\Sigma_{\pm,i}$. Despite this being a trivial cycle in homology we obtain a non-trivial results since $Y'$ is not closed,
	\begin{equation}
		\Delta(\Sigma_{\pm,i})=\frac{3 b_{i+1}}{2}N\, ,
	\end{equation}
	where
	\begin{equation}
		b_{i,\pm}=\frac{1}{2}\left(\Phi_i-\frac{p_i}{n_+n_-}\epsilon\right)\, ,
	\end{equation}
	and\footnote{These are given in \cite{Faedo:2022rqx} with some minor differences in naming conventions and choice of magnetic charges. See also \cite{BenettiGenolini:2023ndb}.} 
	\begin{equation}
		\begin{split}
			\epsilon&=-\frac{n_- n_+(n_--n_+)(n_-+n_++s)}{2(n_--p_1)(n_+-p_1)(s+2(n_-+n_+))}\, ,\\
			\Phi_1&=\frac{p_1(n_-+n_++s)(s+6 p_1-4(n_-+n_+))}{6(n_--p_1)(n_+-p_1)(s+2(n_++n_-))}\, \\
			\Phi_2&=2+\frac{\epsilon(n_+-n_-)}{n_- n_+}-\Phi_1\, .
		\end{split}
	\end{equation}
	This form is motivated by the localisation results in \cite{BenettiGenolini:2023ndb}. Furthermore, one can see that these reduce correctly to the results for the BBBW case in equation (3.35) of \cite{Bah:2013wda} and (3.45) of \cite{BenettiGenolini:2023ndb}. For the analogous M2-branes ending on the $\text{M5}_{\pm}^{\vec{\alpha}_{\pm}}$ there are additional factors of $\vec{\alpha}_{\pm}$ to take into account. We were unable to find a closed form expression for the more general M2-brane, nevertheless we have shown that there is a rich and different structure of probe M2-branes (and therefore BPS operators) in the SCFT arising from compactifying M5-branes on a spindle in comparison to the compactification on a Riemann surface.

	\section{Symmetries and anomalies}                     %
	\label{sec:anomalies}                                  %
	
	After carefully scrutinising the internal geometry and identifying the precise form of the singularities in the eleven-dimensional geometry, we proceed to analyse the global symmetries and 't Hooft anomalies of the dual SCFTs. The eleven-dimensional geometry admits a $\uu(1)^3$ isometry algebra dual to the superconformal $R$ symmetry and two additional $\uu(1)$ flavour symmetries of the dual SCFT. In addition, the dual SCFT enjoys flavour symmetries associated to the $\bC^3/\bZ_{n_\pm}$ orbifold singularities. 
	
	The analysis of the symmetries as well as their 't Hooft anomalies can be performed systematically, following the anomaly inflow methods developed in \cite{Bah:2018gwc,Bah:2018jrv,Bah:2019rgq,Bah:2019vmq,Bah:2020uev}, building on the work of \cite{Harvey:1998bx,Freed:1998tg}. This analysis employs an auxiliary twelve-manifold $\cM_{12}$, realised as an $\cM_6$ fibration over a closed six-manifold $\cN_6$, i.e. $\cM_6 \hookrightarrow \cN_{12}\rightarrow \cN_6$. The space $\cN_6$ is interpreted as a fibration over the external spacetime on which the anomaly polynomial lives. In particular, we will obtain the anomaly polynomial of the dual four-dimensional SCFT by integrating the twelve-form $\cI_{12}$ over the internal space $\cM_6$. 
	\begin{equation}
		\cI_{6} = \int_{\cM_6} \cI_{12}\,.
	\end{equation} 
	Moreover, through the usual descent procedure, the six-or twelve-form anomaly polynomial is related to a five- or eleven-form which can be identified with the topological couplings in the relevant five- or eleven-dimensional supergravity theory,
	\begin{equation}
		\cI_{d+2} = \dd \cI^{(0)}_{d+1}\,, \qquad\qquad S^{\rm (top)}_{d+1} = \int_{\cN_d} \cI^{(0)}_{d+1}\,.
	\end{equation} 
	The twelve-form $\cI_{12}$ was described in \cite{Harvey:1998bx,Freed:1998tg} and is given by
	\begin{equation}\label{eq:defI12}
		\cI_{12} = \f16 E_4 \wedge E_4 \wedge E_4 + E_4 \wedge X_8\,,
	\end{equation}
	where
	\begin{equation}\label{eq:X8M11}
		X_8 = \f1{192}\left[ p_1(T\cM_{11})^2 - 4\, p_2(T\cM_{11}) \right]\,.
	\end{equation}
	The four-form $E_4$ appearing in \eqref{eq:defI12} is a globally defined, closed four-form on the total space $\cN_{12}$ with integral periods on all four-cycles in $\cN_{12}$. When restricted to the fibre over a generic point it reproduces the $G_4$ flux that supports the $\AdS_5$ solution.
	
	\subsection{Construction of \texorpdfstring{$E_4$}{E4}}  %
	
	The four-form $E_4$ can be constructed from the $\UU(1)^3$ equivariant completion of $\wti G_4 = (2\pi\ell_p)^{-3} G_4$, where the prefactor is added to have integer quantisation. The construction of $E_4$ proceeds in two steps, first we define the $\uu(1)^3$ equivariantly closed form $V_4$ such that
	\begin{equation}
		\dd_{\xi_I} V_4 = \left(\dd - \f{\cF^I}{2\pi} \wedge \iota_{\xi_I}\right) V_4 = 0\,,
	\end{equation}
	where we chose a basis $\xi^I = \{ \xi_R, \xi_1, \xi_2\}$ for the Killing vectors and $\cF^I = \dd\cA^I = \{ \cF_R , \cF_1 , \cF_2 \}$ denote the external field strengths associated to the R-symmetry and two flavour symmetries. The Killing vectors for the flavour symmetries can be chosen simply as $\xi_i= \partial_{\phi_i}$. Note however that the Killing vector dual to the R-symmetry is not simply $\partial_z$ but instead $\partial_{\psi}$ where $\psi = -\f32 \fr_z z - \phi_1 - \phi_2$. After putting the metric in the canonical $\cN=1$ form this can be deduced straightforwardly by demanding that the complex three-form $\Omega$ has charge $+1$ under the R-symmetry but is uncharged under the remaining flavour symmetries, see appendix \ref{app:GSMW}.
	
	A suitable ansatz for $V_4$ is given by
	\begin{equation}
		V_4 = \wti G_4 + \sum_I \f{\cF^I}{2\pi} \wedge \omega_{(2)I} + \sum_{I,J} \f{\cF^I}{2\pi} \wedge \f{\cF^J}{2\pi} \,\omega_{(0)IJ}\,.
	\end{equation}
	Demanding equivariant closure then amounts to imposing the following equations
	\begin{equation}\label{eq:equiconstraints}
		\dd \wti G_4 = 0\,,\qquad \iota_{\xi_I} \wti G_4 - \dd \omega_{(2)I} = 0\,, \qquad \iota_{\xi_I} \wti \omega_{(2)J} - \dd \omega_{(0)IJ} = 0\,,
	\end{equation}
	After identifying the appropriate equivariantly closed form $V_4$ we can obtain an expression for $E_4$ by gauging $V_4$ with respect to the isometries. 
	\begin{equation}\label{eq:equiform}
		E_4 = V_4^g =  \wti G_4^g + \sum_I \f{\cF^I}{2\pi} \wedge \omega_{(2)I}^g + \sum_{I,J} \f{\cF^I}{2\pi} \wedge \f{\cF^J}{2\pi} \,\omega_{(0)IJ}\,,
	\end{equation}
	where the superscript $g$ denotes the gauging of the internal forms $\wti G_4$ and $\omega_{(2)I}$. More concretely, this gauging operation consists of replacing in each form on $\cM_6$,
	\begin{equation}
		\dd \psi \rightarrow \cD \psi = \dd \psi - \f{\cA_R}{2\pi}\,, \qquad\qquad \dd\phi_i \rightarrow \cD\phi_i = \dd\phi_i - \f{\cA_i}{2\pi}\,.
	\end{equation}
	It is then straightforward to show that $E_4$ is a closed form on the auxiliary manifold $\cN_{12}$.
	
	Imposing equations \eqref{eq:equiconstraints} uniquely fixes\footnote{Unique up to equivariantly exact pieces but these are irrelevant for our purposes since they do not contribute to any observable.} the internal forms $\omega_{(0)IJ}$ and $\omega_{(2)I}$ to be,
	\begin{equation}\label{eq:equiformses}
		\begin{aligned}
			\omega_{(2)R} =& - \f{\e^{3\lambda}Y^\prime}{3}\,, &
			\omega_{(0)RR} =& \f19 y\,, \\
			\omega_{(2)\phi_1} =& -\iota_{\phi_1}C_3 + \f19 \dd y \wedge \dd \psi \,,&
			\omega_{(0)R\phi_1} =& -\f y{6}\f{w}{h_2(w)} \,,\\
			\omega_{(2)\phi_2} =& -\iota_{\phi_2}C_3 + \f19 \dd y \wedge \dd \psi\,,&
			\omega_{(0)R\phi_2} =& -\f y{6}\f{w}{h_1(w)}\,,\\
			\omega_{(0)\phi_1\phi_1} =& 0 \,,&
			\omega_{(0)\phi_1\phi_2} =& 0\,,\\
			\omega_{(0)\phi_2\phi_2} =& 0\,.&&
		\end{aligned}
	\end{equation}
	Inserting these expressions in \eqref{eq:equiform} we have provided an explicit representative of the globally well-defined form $E_4$ satisfying all the requirements listed above. In the above expressions, the two-form $Y^\prime$ is the calibration form, introduced in Appendix \ref{app:GSMW}. Note that the zero- and two-form part associated to the R-symmetry are identical to the expressions found in \cite{BenettiGenolini:2023kxp,BenettiGenolini:2023yfe}.
	
	\subsection{Flavour symmetries from punctures}   %
	
	In addition to the flavour symmetries discussed so far, arising from isometries of the internal manifold, the dual SCFT can have baryonic zero-form flavour symmetries arising from non-trivial two-cycles in the internal geometry. Integrating the M-theory three-form on such cycles gives rise to vector potentials in five dimensional gauged supergravity sitting in so-called Betti multiplets. More generally, wrapping the M-theory three-form on non-trivial $p$-cycles in the internal manifold can give rise to $(2-p)$-form symmetries in the dual SCFT. In our setup, such additional non-trivial cycles arise from the blow-up of the $\bC^3/\bZ_{n_\pm}$ orbifold singularities. 
	
	In order to understand the resulting flavour symmetry we first will have a look at the resolutions of generic $\bC^3/\Gamma$ orbifolds, where $\Gamma$ is a finite subgroup of $\SU(3)$, following the work of Ito and Reid \cite{Ito-Reid}. In our context, the requirement that $\Gamma\subset \SU(3)$ can be physically motivated by the fact that this class of orbifolds is Calabi-Yau and therefore gives rise to at least $\cN=1$ supersymmetry in the dual SCFT. Indeed, this requirement has already been derived in seven dimensions when demanding the existence of a globally well-defined Killing spinor.
	
	Denote the conjugacy classes of $\Gamma$ by $\gamma_i$. All the group elements in a given conjugacy class have the same eigenvalues $\lambda_i$, which can be written as
	\begin{equation}
		\lambda_i = \left\{ \exp\left(2\pi \ii\,\f{a_i}{r_i}\right)\,,\, \exp\left(2\pi \ii\,\f{b_i}{r_i}\right)\,,\, \exp\left(2\pi \ii\,\f{c_i}{r_i}\right)\right\}\,,
	\end{equation}
	where $r_i$ is the order of the elements in $\gamma_i$ and $0\leq a_i,b_i,c_i \leq r_i$. Define the age of the conjugacy class as the integer ${\rm age}(\gamma_i) = \f{1}{r_i}(a_i+b_i+c_i)$ and denote the set of conjugacy classes of age $m$ by $\Gamma_m$. Ito and Reid proved that for each orbifold $\bC^3/\Gamma$ there exists a crepant resolutions
	\begin{equation}
		\pi : \wti X \rightarrow \bC^3/\bZ_k\,,    
	\end{equation}
	such that $b_{2m}(\wti X) = |\Gamma_m|$ and $b_3(\wti X) = 0$. 
	
	Coming back to our setup, the exceptional divisors $S_\alpha \in H_4(\wti X,\bR)$ from the blow-up are dual to the compactly supported 2-forms $\wti\omega_{(2)\alpha} \in H^2_c(\wti X,\bR)$ which can be thought of as forms localised at the singular points in our original geometry. We can therefore consider the following additional terms in the expansion of $E_4$. 
	\begin{equation}
		E^{\rm punc.}_4 = \sum_{\alpha=1}^{b_{4}(\wti X)} \f{\cF^\alpha}{2\pi} \wedge \wti\omega_{(2)\alpha}\,.
	\end{equation}
	where $\cF^\alpha = \dd \cA^\alpha$ are external gauge fields for the resulting flavour symmetries. In the lower dimensional gauged supergravity these therefore give rise to massless gauge fields in the Cartan of the gauge group, where the rank of the gauge group is given by $r=b_4(\wti X)$. In various cases, the gauge symmetry gets enhanced to a non-abelian symmetry at the SCFT point. 
	
	In particular, one can wrap M2-branes on rational curves inside a compact divisor $S_\beta$. These curves can be described by a geometric ruling of the divisor by rational curves $f_\beta$, i.e. $\bC\bP^1$s. In the dual QFT (in the Coulomb phase) such wrapped M2-branes give rise to charged particles, i.e. W-bosons. The charge of the particle under the Cartan corresponding to a divisor $S_\alpha$ is given by
	\begin{equation}
		S_\alpha \cdot f_\beta  = - C_{\alpha\beta}\,,
	\end{equation}
	where the left hand side denotes the intersection number in $\wti X$ and can be computed from the triple intersection numbers $S_\beta^2 \cdot S^\alpha$. The mass of the particle is determined by the symplectic volume of $f_\alpha$ such that in the singular limit where the resolution cycles shrink these particles become the massless W-bosons enhancing the $\uu(1)^r$ symmetry to a possibly non-abelian $\fg$ symmetry. In particular, the matrix $C_{\alpha\beta}$ reproduces the entries of Cartan matrix of the flavour symmetry of the dual SCFT. In addition to providing a holographic description of the flavour symmetry associated to the punctures the additional terms in the expansion of $E_4$ give rise to Chern-Simons terms in the five-dimensional supergravity theory which describe the (mixed) 't Hooft anomalies of the dual SCFT which are therefore controlled by the triple intersection numbers of compact divisors $k_{\alpha\beta\gamma} = S_\alpha \cdot S_\beta \cdot S_\gamma$. Furthermore, the crepant resolutions do not have odd dimensional cycles, so they do not contribute to the one-form symmetry of the dual SCFTs. 
	
	This setup is very similar and was heavily inspired by geometric engineering of 5d S(C)FTs in M-theory, see \cite{Morrison:1996xf,Douglas:1996xp,Intriligator:1997pq} and in particular \cite{Xie:2017pfl,Apruzzi:2019kgb,Closset:2020scj,Closset:2020afy,Eckhard:2020jyr,Tian:2021cif,Closset:2021lwy,Mu:2023uws} for the applications with orbifolds and more general isolated canonical singularities. However, the crucial difference is that in our case the non-compact Calabi-Yau three-fold only gives a local description of a globally compact three-fold. Hence, the 5d theory obtained after compactification is a supergravity theory. For this reason we are solely interested in the compact cycles resulting in gauge symmetries in the bulk. In particular we ignore the effect of the additional $b_2(\wti X) - b_4(\wti X)$ compact two-forms dual to non-compact divisors which in geometric engineering encode (part of) the flavour symmetry of the 5d S(C)FT. 
	
	So far our discussion has been fairly general and applies to any finite subgroup of $\SU(3)$. The orbifolds we encounter in the uplift of our seven-dimensional solutions however are all abelian orbifolds of the form $\bC^3/\bZ_n$. From now on we specify to this case. Note that for these cases all conjugacy classes contain a single element such that $n=1+|\Gamma_1|+|\Gamma_2| = 1+b_2(\wti X)+b_4(\wti X)$ and furthermore their resolution is always toric. This class of punctures is surprisingly rich and depending on the `charges' $m^{(i)}_\pm$ a variety of orbifold actions is possible. For example, for $\bZ_6$ the generator can be chosen either as $\f16(1,1,4)$ or $\f16(1,2,3)$.\footnote{We only include those actions resulting in a three-dimensional singularity. When the orbifold action acts trivially in one of the directions we recover the usual $\cN=2$ punctures.} Note that in these two cases the fourth Betti number differs so that the rank of the flavour groups is respectively $2$ or $1$. 
	
\begin{figure}[!htb]
		\centering
		\includegraphics[scale=0.7]{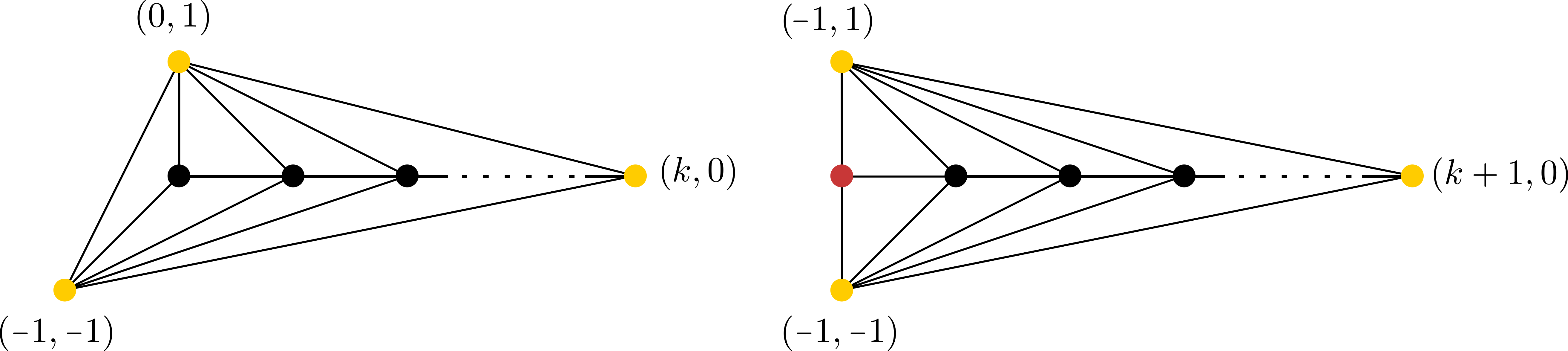}
		\caption{The toric diagrams for the two infinite families of $\bZ_n$ singularities introduces above. The black dots indicate compact divisors, while the yellow ones denote non-compact divisors. The red dot denotes the compact divisor corresponding to a flavour curve in the geometric engineering setup. The toric diagram on the left corresponds to $n=2k+1$, while the diagram on the right corresponds to $n=2k+2$. Note that since these are toric Calabi-Yau threefolds they can be encoded in a two-dimensional diagram where we fixed the third coordinate of all rays to be one.} 
		\label{fig:toricdiag}
	\end{figure}

	Before we continue let us briefly consider a few examples. A first infinite class of examples is given by $\Gamma=\bZ_n$ with generator $\f1n (1,1,n-2)$. For $n$ odd res. even, the resolution of this singularity can be encoded in the toric diagrams in Figure \ref{fig:toricdiag}. Computing the intersection numbers corresponding to the Cartan matrix and Chern-Simons coefficients now becomes a straightforward exercise in toric geometry. For a review of toric geometry we refer the reader to one of the excellent textbooks on the subject.
	
	Up to $\SL(2,\bZ)$ transformations the toric diagram is obtained from the three vertices of weighted projective space, $\bC\bP^{p,q,r}$ by interpreting them as the rays of the fan of the toric three-fold singularity.\footnote{A set of vertices for weighted projective space can be obtained by finding (coprime) integer solutions to the equation $p v_1 + q v_2 + r v_3 = 0$.} After an $\SL(2,\bZ)$ transformation the $\Gamma=\bZ_n$ with generator $\f1n (1,1,n-2)$ then takes the form as in Figure \ref{fig:toricdiag}. From this toric diagram we can then read off the gauge symmetry in $\AdS_5$ or equivalently the flavour symmetry of the dual SCFT. There are $k$ compact divisors, which correspond to the Cartan of the flavour algebra. Computing the Cartan matrix from the intersection numbers we find that the full non-abelian flavour symmetry of the dual SCFT is given by $\SU(k)$ where $n=2k+1$ or $n=2k+2$. Note that a single orbifold point in seven dimensions get lifted to a pair of orbifold singularities in eleven dimensions so that the total global symmetry of the puncture is given by $\SU(k)^2$. 
	
	The conical singularity of the spindle gets doubled in the internal space which explains the doubling of the flavour symmetry. Indeed, notice that the M2 branes wrapping $\Sigma_{\pm,i}$, extending from a singularity in the internal space to the probe M5-brane on the diagonal face in \ref{fig:prism}, are charged under one of these flavour symmetries. This provides evidence that indeed this $\cN=1$ puncture gives rise to the doubled $\SU(k)^2$ flavour symmetry.
	
	More explicitly, in the case of odd $n=2k+1$ the compact surfaces are given by $\bC\bP^2\cup \mathbb{F}_3\cup \mathbb{F}_5\cup \cdots \cup \mathbb{F}_{2k-1}$ and the triple intersection numbers are as follows. The triple intersection of three non-identical compact divisors always vanishes. The intersection numbers or two identical and one distinct neighbouring divisor are given by
	\begin{equation}
		\begin{aligned}
			S_0^2 \cdot S_1 &= 1\,,\quad & S_1^2 \cdot S_0 &= -3\,,\quad & S_1^2 \cdot S_2 &= 3\,,\quad & S_2^2 \cdot S_1 &= -5 \,,\\
			S_2^2 \cdot S_3 &= 5\,, \quad & &\cdots \quad & S_{k-2}^2\cdot S_{k-1} &= 2k-3\,,\quad & S_{k-1}^2\cdot S_{k-2} &= -2k+1\,.    
		\end{aligned}
	\end{equation}
	Finally when the three divisors are identical, the triple intersection is $S_0^3 = 9$ and $S_i^3 = 8$, where $i=1,\cdots,k-1$.

\begin{figure}[!htb]
		\centering
		\includegraphics[scale=0.7]{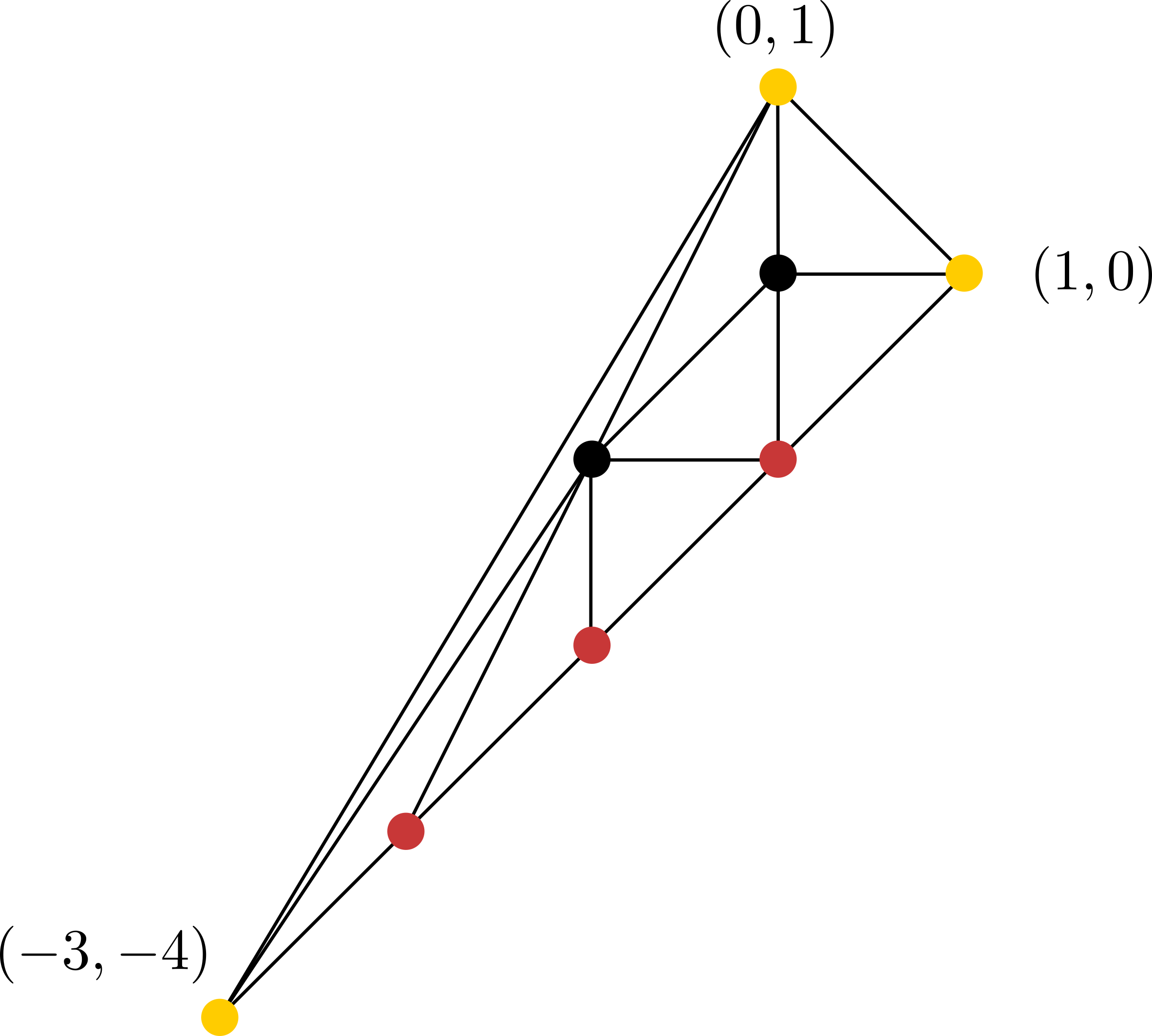}
		\caption{The toric diagrams for the $\bZ_8$ singularity with generator $\f18(1,3,4)$. The black dots indicate compact divisors, while the yellow ones denote non-compact divisors. The red dots denote the compact divisors corresponding to flavour curves in the geometric engineering setup.} 
		\label{fig:toricdiag8}
	\end{figure}

	Similarly, as a last example we consider the singularity $\Gamma = \bC^3/\bZ_8$ with generator $\f{1}{8}\left( 1 , 3 , 4 \right)$. The toric diagram is given in Figure \ref{fig:toricdiag8} from which we can read of the rank of the flavour symmetry group which in this case is two. However, in this case we do not find a $\SU(3)$ flavour symmetry but instead find that the flavour symmetry is $\SU(2)\times \SU(2)$. Finally, in Table \ref{tab:rankGamma} we collect the Betti numbers and flavour symmetries of the dual SCFTs for a variety of abelian orbifolds of order $n<14$. \footnote{This table is modified from Table 5 in \cite{Tian:2021cif}.} 

	\begin{table}[!htb]
		\centering
		\begin{tabular}{c|c|c|c|c}
			$n$ & $\quad$Generator$\quad$ & rank $=\,b_4(\wti X)$ & $\quad b_2(\wti X)\quad$ & Flavour algebra \\
			\hline\hline
			$3$ & $\TBstrut\f13(1,1,1)$ & $1$ & $1$ & --- \\
			$4$ & $\TBstrut\f14(1,1,2)$ & $1$ & $2$ & $\su(2)$ \\
			$5$ & $\TBstrut\f15(1,1,3)$ & $2$ & $2$ & --- \\
			$6$ & $\TBstrut\f16(1,1,4)$ & $2$ & $4$ & $\su(2)$  \\
			$6$ & $\TBstrut\f16(1,2,3)$ & $1$ & $4$ & $\su(2)$  \\
			$7$ & $\TBstrut\f17(1,1,5)$ & $3$ & $3$ & --- \\
			$7$ & $\TBstrut\f17(1,2,4)$ & $3$ & $3$ & --- \\
			$8$ & $\TBstrut\f18(1,1,6)$ & $3$ & $4$ & $\su(4)$  \\
			$8$ & $\TBstrut\f18(1,2,5)$ & $3$ & $4$ & --- \\
			$8$ & $\TBstrut\f18(1,3,4)$ & $2$ & $5$ & $\su(2)\times \su(2)$  \\
			$9$ & $\TBstrut\f19(1,1,7)$ & $4$ & $4$ & --- \\
			$9$ & $\TBstrut\f19(1,2,6)$ & $3$ & $5$ & --- \\
			$10$ & $\TBstrut\f1{10}(1,1,8)$ & $4$ & $5$ & $\su(5)$  \\
			$10$ & $\TBstrut\f1{10}(1,2,7)$ & $4$ & $5$ & --- \\
			$10$ & $\TBstrut\f1{10}(1,3,6)$ & $4$ & $5$ & --- \\
			$10$ & $\TBstrut\f1{10}(1,4,5)$ & $2$ & $7$ & $\su(2)\times \su(2)$  \\
			$11$ & $\TBstrut\f1{11}(1,1,9)$ & $5$ & $5$ & --- \\
			$11$ & $\TBstrut\f1{11}(1,2,8)$ & $5$ & $5$ & --- \\
			$12$ & $\TBstrut\f1{12}(1,1,10)$ & $5$ & $6$ & $\su(6)$  \\
			$12$ & $\TBstrut\f1{12}(1,2,9)$ & $4$ & $7$ & $\su(4)\times \su(2)$ \\
			$12$ & $\TBstrut\f1{12}(1,3,8)$ & $3$ & $8$ & $\su(3)\times\su(2)$ \\
			$12$ & $\TBstrut\f1{12}(1,4,7)$ & $4$ & $7$ & --- \\
			$12$ & $\TBstrut\f1{12}(1,5,6)$ & $3$ & $8$ & $\quad \su(2)\times \su(2)\times \su(2)\quad$ \\
			$13$ & $\TBstrut\f1{13}(1,1,11)$ & $6$ & $6$ & --- \\
			$13$ & $\TBstrut\f1{13}(1,2,10)$ & $6$ & $6$ & --- \\
			$13$ & $\TBstrut\f1{13}(1,3,9)$ & $6$ & $6$ & --- \\
		\end{tabular}
		\caption{List of (inequivalent) $\bC^3/\bZ_n$ orbifolds with their generators and the Betti numbers of the (crepant) resolution. The last column gives the precise form of the flavour symmetry algebra (if known) which can be obtained using the methods of \cite{Tian:2021cif}. The dashes `---' denote the cases where the methods above do not predict symmetry enhancement to a non-abelian flavour symmetry and we have $\uu(1)^\text{rank} \subseteq \fg_F$.}
		\label{tab:rankGamma}
	\end{table}

	The uplift of the seven-dimensional solutions shown in this paper necessarily only provide a subset of the possible solutions in eleven dimensions. Therefore it is reasonable to expect that after uplifting to eleven dimensions one can significantly generalise the type of punctures. In particular, a natural expectation is that more generic finite subgroups of $\SU(3)$ such as $\bC^3/\bZ_n\times \bZ_m$ singularities or even non-abelian $\Gamma$ can arise as $\cN=1$ punctures in class $\cS$. It would be very interesting to find a local description of such punctures by carefully analysing the associated Monge-Amp\'ere equations analogous to the Toda equation (or cylindrical Laplacian) in the case of $\cN=2$ punctures \cite{Gaiotto:2009gz}. One can then use similar tools as above to analyse the flavour symmetry originating from the puncture in this more general setup. Note however that not all of the more general puncture are toric. In such case more general methods, see for example \cite{Tian:2021cif}, are needed to resolve the singularities. 
	
	\subsection{Anomaly inflow}    %
	
	Having constructed the equivariantly closed form $E_4$ and analysed the additional flavour symmetries from the punctures, it is now an algorithmic task to compute $\cI_{12}$ and integrate along $\cM_6$. Writing $E_4$ as,
	\begin{equation}
		E_4 = G_4^{(g)} +\f{\cF^a}{2\pi} \wedge \omega_{(2)a}^{(g)}  + \f{\cF^a}{2\pi} \wedge \f{\cF^b}{2\pi}\,\omega_{(0)ab}\,, 
	\end{equation}
	where the index $a,b \in \{ I\,, \alpha \}$ collectively run over the flavour symmetries from isometries as well as two-cycles in the internal manifold. Substituting this into the expression for $\cI_{12}$ we find that the $E_4^3$ term becomes,
	\begin{equation}
		\f16 E_4^3 = \f{\cF^a}{2\pi}\wedge \f{\cF^b}{2\pi} \wedge \f{\cF^c}{2\pi} \wedge\left( \wti G_4 \wedge \omega_{(2)a} \omega_{(0)bc} + \f16\omega_{(2)a}\wedge\omega_{(2)b}\wedge\omega_{(2)c}   \right) + \cdots \,,
	\end{equation}
	where the dots denote terms that do not have the appropriate number of legs along the internal space and hence will not contribute to the anomaly polynomial of the resulting four-dimensional theory. Similarly, the higher derivative contributions from $E_4\wedge X_8$ give rise to the following contributions,
	\begin{equation}
		E_4\wedge X_8 = -\f{1}{96} \f{\cF^a}{2\pi} \wedge p_1(T \AdS_5) \wedge \omega_{(2)a}\wedge p_1(T\cM_6) + \cdots \,,
	\end{equation}
	where we used the splitting principle to rewrite \eqref{eq:X8M11} in terms of the Pontryagin classes of the internal resp. $\AdS$ space. In order to obtain the anomaly polynomial of the resulting four-dimensional theory all that is left is to integrate these expressions over the internal space $\cM_6$. The resulting expression is given by
	\begin{equation}
		\cI_6 = \f16 k_{abc}\, c_1(\cF^a)\wedge c_1(\cF^b) \wedge c_1(\cF^c) - \f1{24} k_a\, c_1(\cF^a) \wedge p_1(T\AdS_5)\,,
	\end{equation}
	where $k_{abc}$ and $k_a$ are defined as
	\begin{align}
		k_{abc} =& \int_{\cM_6} \left(  \omega_{(2)a}\wedge \omega_{(2)b}\wedge \omega_{(2)c} + 6 \wti G_4 \wedge \omega_{(2)a} \omega_{(0)bc} \right)\,, \label{eq:kabc}\\
		k_a =&  \f{1}{4} \int_{\cM_6}  \omega_{(2)a}\wedge p_1(T\cM_6) \label{eq:ka} \,,
	\end{align}
	and the terms proportional to $\wti G_4$ vanish identically. Comparing this with the anomaly polynomial of a generic four-dimensional $\cN=1$ SCFT,
	\begin{equation}
		\cI_6 = \f{8}{27}\left(5a-3c\right) c_1(\cF_R)^3 + \f23 \left( c- a \right) c_1(\cF_R) p_1\left(T\cM_4 \right)\,,
	\end{equation}
	the above expressions reduce to the well-known expressions for the conformal anomalies $a$ and $c$,
	\begin{equation}
		a = \f3{32}\int_{\cM_6} \left(  3\,k_{RRR} - k_R \right)\,,\qquad\quad c = \f1{32}\int_{\cM_6} \left( 9 \,k_{RRR} +5\, k_R \right)\,.
	\end{equation}
	At leading order in the large $N$ expansion we have $a\simeq c$ and it follows from \eqref{eq:equiformses} above that our expression reduces precisely to
	\begin{equation}
		a \simeq \frac{N^3}{128\pi^3}\int_{\cM_6} \e^{9\lambda} \,\dvol_{\cM_6}\,,
	\end{equation}
	which agrees with the expression obtained in \cite{Gauntlett:2006ai}. Note that with the choice of $E_4$ introduced above one obtains the integrand up to total derivatives only. The end result after integration is obviously independent of such terms. We can make this more explicit by explicitly performing the equivariant integration of $E_4^3$ over $\cM_6$. Indeed, in line with the results in \cite{BenettiGenolini:2023kxp,BenettiGenolini:2023ndb} we find that the leading order central charge is given by
	\begin{equation}
		a = \f{N^3}{2^5 3^4}\sum_{F_0} \f{1}{d_{F_0}}\f{y^3}{\epsilon_1\epsilon_2\epsilon_3}\,.
	\end{equation}
	This sum goes over the four fixed orbifold points where the weights are given in \eqref{eq:orbweights}. Hence, the resulting expression for the conformal $a$-anomaly manifestly agrees with \cite{Ferrero:2021wvk,BenettiGenolini:2023kxp,BenettiGenolini:2023ndb}. At subleading order in $N$, $a$ and $c$ typically differ. From the anomaly polynomial above we can read off that
	\begin{equation}
		a-c = \f{N}{128\pi} \int_{\cM_6} \omega_{(2)R}\wedge p_1 (T\cM_6)\,,
	\end{equation}
	which can be computed by direct integration or more elegantly using equivariant localisation. The flavour central charges $k_{abc}$ resulting from the isometries of the internal manifold can be computed by direct (equivariant) integration after substituting the relevant forms from \eqref{eq:equiformses}. The $k_{\phi_i\phi_j\phi_k}$ are trivially vanishing. In addition, in line with $a$-maximisation we find that $\partial_s a(s) \simeq k_{RR\phi_i} = 0$ and $\partial^2_s a(s) \simeq k_{R \phi_i \phi_j}<0$. The computation of the flavour central charges for the flavours corresponding to the puncture can be computed as the triple intersection number 
	\begin{equation}
		k_{\alpha\beta\gamma} = S_\alpha\cdot S_\beta \cdot S_\gamma\,,
	\end{equation}
	which can be read off from the toric diagram as instructed above. Finally, we can consider the mixed flavour central charges between the R-symmetry or flavour symmetries from isometries and the flavour symmetries from the puncture. Since the puncture divisors are localised at the puncture location and provide a basis of harmonic two-forms one should expand the two form corresponding to the R-symmetry and mesonic flavour symmetries in terms of the localised cycles at the puncture location. After this one can compute these mixed flavour central charges in a similar manner as a linear combination of intersection numbers. However, we do not pursue this computation in the present work. Similarly, we leave the computation of the $k_a$ open for future research endeavours. Note however that, as can be observed from \cite{Ferrero:2021wvk,BenettiGenolini:2023kxp,BenettiGenolini:2023ndb}, in all anomalies and central charges related to the internal symmetries, the dependence on $m_\pm^{(i)}$ can be absorbed in the fluxes $p_i$. On the other hand, the flavour anomalies for the additional global symmetries arising from the four orbifold singularities do contain dependence on the $m_\pm^{(i)}$. This is natural as we expect these anomalies to explicitly depend on the specific choice of punctures.

	
	\bigskip
	\bigskip
	\bigskip
	
	\leftline{\bf Acknowledgements}
	\smallskip
	\noindent It is with pleasure that we thank Ibou Bah, Sakura Schafer-Nameki and James Sparks for useful and inspiring discussions. The contributions of PB were made possible through the support of grant No. 494786 from the Simons Foundation (Simons Collaboration on the Non-perturbative Bootstrap) and the ERC Consolidator Grant No. 864828, titled “Algebraic Foundations of Supersymmetric Quantum Field Theory'' (SCFTAlg). CC wishes to express gratitude for the support received from the Mathematical Institute of the University of Oxford.

	\newpage                                                             %
	\appendix                                                            %
	
	\section{Uplift formulae}    %
	\label{app:upliftformulae}   %
	
	The general formulae to uplift a solution of maximal 7d gauged supergravity were derived in \cite{Nastase:1999kf,Nastase:1999cb,Cvetic:1999xp} which, for completeness, we restate here. The eleven-dimensional metric is given by
	\begin{equation}
		L^{-2}\ds_{11}^2 = \Delta^{1/3}\, \ds_7^2 + \Delta^{-2/3}\left( X_0^{-1}\,\dd\mu_0^2 + \sum_{i=1}^{2}X_i^{-1}\left( \dd\mu_i^2 + \mu_i^2 \,D\phi_i^2 \right) \right)\,,
	\end{equation}
	where we defined the function
	\begin{equation}
		\Delta = \sum_{I=0}^{2} X_I \,\mu_I^2\,,
	\end{equation}
	in terms of the seven-dimensional supergravity scalars $X_I$ where $X_0 = X_1^{-2}X_2^{-2}$. The coordinates $\phi_i$ are $2\pi$-periodic, while the embedding coordinates $\mu_I$ satisfy $\mu_0^2+\mu_1^2+\mu_2^2 = 1$. A convenient choice for our purposes is, 
	\begin{equation}
		\mu_0 = \cos\xi\,,\qquad \mu_1 = \sin\xi\cos\theta\,,\qquad  \mu_2 = \sin\xi\sin\theta\,.
	\end{equation}
	where $\xi\in[0,\pi]$ and $\theta\in [0,\pi/2]$. In addition, we defined the gauged one-forms
	\begin{equation}
		\DD\phi_i = \dd\phi_i - A^{(i)}\,.
	\end{equation}
	The eleven-dimensional three-form potential in turn can be written as,
	\begin{equation}
		\begin{aligned}
			L^{-3}C_3 =& \,\left[ -\frac{1}{4\mu_0}\left(\dd \mu_1^2-\dd \mu_2^2\right) + \frac{\mu_1\,\mu_2}{\Delta\,\mu_0}\left( X_2\,\mu_2 \,\dd \mu_1- X_1\,\mu_2\,\dd\mu_2\right)\right]\wedge \DD\phi_1\wedge \DD\phi_2\\
			&\,\,\,\,-\frac{\mu_0}{2} \left(F^{(1)}\wedge \DD\phi_2+F^{(2)}\wedge \DD\phi_1\right)\,.
		\end{aligned}
	\end{equation}
	%

	\section{Canonical \texorpdfstring{$\cN=1$}{N=1} form}      %
	\label{app:GSMW}                                            %

	In this appendix we introduce the relevant definitions and coordinate changes in order to transform our solutions into the canonical form for $\cN=1$ $\AdS_5$ solutions of M-theory \cite{Gauntlett:2004zh}, henceforth referred to as the GMSW form. Since we know the spinors of the seven-dimensional solution from \cite{Ferrero:2021etw} we can work directly in the six-dimensional internal space $\cM_6$ and to construct the eleven-dimensional Killing spinor. From this one can then construct all the relevant bilinear and reconstruct the full eleven-dimensional background.
	
	Although conceptually the above approach is more straightforward, here we take a slightly roundabout way to obtain the transformation into the canonical form which turns out to be computationally easier. We use that there is a general truncation of the GMSW classification to five-dimensional minimal gauged supergravity as given in \cite{Gauntlett:2006ai}. On the other hand, in \cite{Cheung:2022ilc} it was shown how to perform the truncation of seven-dimensional U$(1)^2$ gauged supergravity on a spindle to five-dimensional minimal gauged supergravity. By using the two truncations from 11d, which must agree, we are able to extract out the embedding of the spindle solution into the GMSW form.
	
	The metric for most general $\mathcal{N}=1$ supersymmetric $\AdS_5$ solution can be written in the form
	\begin{equation}
		\ds_{11}^2 = \e^{2\lambda} \left[ \ds_{\AdS_5}^2 + \ds_{\cM_6}^2 \right] \,,
	\end{equation}
	where the internal metric is 
	\begin{equation}
		\ds_{\cM_6}^2 = \e^{-6\lambda}g_{ij}(x,y)\dd x^i \dd x^j + \e^{-6\lambda} \sec^2\zeta (x,y) \dd y^2 +\f19 \cos^2\zeta(x,y) (\dd\psi+\rho)^2\,.
	\end{equation}
	Here $g_{ij}$ is the metric on a four-dimensional K\"ahler manifold (at fixed $y$) and $\partial_\psi$ is a Killing vector of the metric $g$, the functions, $\lambda$ and $\zeta$ and the four-form flux $G_4$. This Killing vector realises the R-symmetry in the dual SCFT. The four-form field strength can be written as
	\begin{multline}
		G_4 = -\left( \partial_y \e^{-6\lambda} \right) \dvol_4 - \e^{-9\lambda} \sec\zeta \left( \star_4 \dd_4 \e^{6\lambda}\right) \wedge K_1 - \f13 \cos^3\zeta (\star_4 \partial_y \rho)\wedge K_2\\
		+\e^{3\lambda}\left( \f13\cos^2\zeta \star_4\dd_4\rho - 4\e^{-6\lambda} J \right)\wedge K_1 \wedge K_2\,.
	\end{multline}
	In this expression, $\star_4$, $\dd_4$ and $J$ denote the Hodge star, exterior derivative and K\"ahler two-form with respect to the four-dimensional metric $g_{ij}$. The one-forms $K_1$ and $K_2$ are defined to be
	\begin{equation}
		K_1 = \e^{-3\lambda} \sec\zeta \dd y\,,\qquad K_2 = \f13 \cos \zeta (\dd \psi + \rho) \,,
	\end{equation}
	and $2y=\sin\zeta \e^{3\lambda}$.
	
	We can now uplift the solution in equation \eqref{eq:7dM5sol} to 11d using the results of appendix \ref{app:upliftformulae} and then rewrite into the above form. We need to pick a gauge for the gauge fields in the uplift and so we take the gauge in \eqref{eq:7dM5sol} in the following, though any gauge transformation can be performed by
	\begin{equation}
		A_i\rightarrow A_i+\dd \Lambda_i \quad \Rightarrow \quad \dd\phi_i\rightarrow \dd\phi_i-\dd\Lambda_i\, .
	\end{equation}
	The choice of gauge is important in the main text when analysing the orbifold action, however for the purposes of this appendix it is largely irrelevant. 
	From the form of the eleven-dimensional metric we can immediately read off the form of the warp factor,
	\begin{equation}
		\e^{2\lambda}= \Upsilon_H^{1/3}\, .
	\end{equation}
	Next, after writing out the metric and comparing with the truncation of \cite{Cheung:2022ilc} we can extract out the $y$ coordinate, finding
	\begin{equation}
		y= w \mu_0=w\cos\xi\, ,
	\end{equation}
	and it immediately follows that the one-form dual to the R-symmetry vector is 
	\begin{equation}
		\dd \psi+\rho =-\frac{3}{2}\left(\fr_z\dd z +\frac{1}{\Upsilon_F} \big(\mu_1^2 \dd\phi_1+\mu_2^2\dd\phi_2\big)\right)\, .
	\end{equation}
	Moreover it is simple to show that 
	\begin{equation}
		\cos^2\zeta= \frac{1}{4 w^2}\frac{\Upsilon_H}{\Upsilon_F}\, .
	\end{equation}
	Having extracted these crucial pieces we can identify the two-form $J$ to be 
	\begin{equation}
		\begin{split}
			J=e^1\wedge e^2+e^3 \wedge e^4\, ,
		\end{split}
	\end{equation}
	where the vielbein on the four-dimensional space are 
	\begingroup
	\allowdisplaybreaks
	\begin{align}\label{eq:vielbein4}
		e^1&=\sqrt{ \Theta(w,\theta)} \left(\sin\xi \dd\theta -\frac{\cos\theta\sin\theta\cos\xi (h_1(w)-h_2(w))}{\Theta(w,\theta)}\dd\xi\right)\nn\\
		e^2&= \frac{\cos\theta\sin\theta \sin\xi}{\sqrt{ \Theta(w,\theta)}}\left(h_1(w)\dd\phi_1-h_2(w)\dd\phi_2\right)\, ,\\
		e^3&= \frac{\sqrt{\Theta(w,\theta)}\sin\xi\sqrt{f(w)}}{4 w}\sqrt{\frac{\Upsilon_H}{\Upsilon_F}}\left(\frac{w }{f(w)}\dd w+\frac{4 \cos\xi}{\Theta(w,\theta)\sin\xi }\dd\xi\right)\, ,\nn\\
		e^4&=-\frac{\sin\xi \sqrt{f(w)}}{w\sqrt{\Theta(w,\theta)}}\sqrt{\frac{\Upsilon_H}{\Upsilon_F}}\left(\cos^2\theta\dd\phi_1+\sin^2\theta\dd\phi_2\right)\, .\nn
	\end{align}
	\endgroup
	For notational brevity we defined the function
	\begin{equation}
		\begin{split}
			\Theta(w,\theta)&=\sin^2\theta h_1(w)+\cos^2\theta h_2(w)\, .
		\end{split}
	\end{equation}
	With the above definitions at hand we can compute everything we need. However, note that the vielbein \eqref{eq:vielbein4} naively seems to have legs along five directions in the internal space. This is an artefact of our original choice of coordinates and one can check that with this definition of the vielbein all algebraic and differential constraints of \cite{Gauntlett:2004zh} are satisfied.
	
	In principle one should be able to find a general coordinate change to make this explicit but in practice finding such coordinate transformations is incredibly complicated. However, when $h_1(w)=h_2(w)$ we can make the following change of coordinates,\footnote{This simplification is not possible for the spindle solution where they can never be equal, however it is possible for other global completions. }
	\begin{equation}
		y= w \cos\xi\, ,\quad x_1=\log \tan\theta\, ,\quad x_2=W(w)+4\log\sin\xi\, ,
	\end{equation}
	where
	\begin{equation}\label{eq:Weq}
		W'(w)=\frac{w h_1(w)}{f(w)}\, .
	\end{equation}
	In these coordinates the bracketed part of the two vielbeins $e^1$ and $e^3$ become respectively $\dd x_1$ and $\dd x_2$. Solving \eqref{eq:Weq} one can find an explicit expression for $W(w)$,
	\begin{equation}
		W(w)= \sum_{i=1}^{4}\frac{w_i(q_1+w_i^2)}{2f'(w_i)}\log(w-w_i)\, ,
	\end{equation}
	where the sum goes over the four roots of the function $f(w)$. For $q_1\neq q_2$ one can find an equivalent set of coordinates by expanding in $\epsilon= q_2-q_1$ and perturbatively find a coordinate change. However, the resulting expressions are rather messy and not very illuminating so we refrain from stating them here. Writing the above vielbein with explicit $y$ dependence would be useful to better understand the underlying four-dimensional metric and therefore how one may go about further generalising these solutions. 
	
	Both in the computation of the operator dimensions form wrapped M2 branes, as well as in the computation of the equivariant completion of the form $E_4$ we encountered the calibration form introduced in \cite{Gauntlett:2006ai},
	\begin{equation}\label{eq:Yprime}
		Y^\prime = -\sin\zeta \e^{-6\lambda} J + K_1\wedge K_2\,.
	\end{equation}
	We have also stated that the correct $R$-symmetry is a linear combination of the spindle coordinate and the $S^4$ coordinates. To show this it is useful to consider the holomorphic $(2,0)$ form on the four-dimensional base. One finds:
	\begin{equation}
		\Omega=\e^{-\ii\left(\tfrac{3}{2}\fr_z z+\phi_1+\phi_2\right)}(e^1+\ii e^2)\wedge (e^3+\ii e^4)\, .
	\end{equation}
	From GMSW we read off that the R-symmetry coordinate is given by
	\begin{equation}
		-\psi=\tfrac{3}{2}\fr_z z+\phi_1+\phi_2\, ,
	\end{equation}
	as stated in the main text.

	\section{From spindles to BBBW}     %
	\label{app:BBBW}                    %
	
	In this final appendix we show that the spindle solutions presented in this note in fact include all of the solutions studied in BBBW \cite{Bah:2011vv,Bah:2012dg}. For genus $0$ the relevant limit is rather straightforward and was already presented in \cite{Ferrero:2021etw}. The goal of this appendix is to explicitly demonstrate that more general limits recover the BBBW solution for arbitrary genus Riemann surfaces. 
	
	Recall that the metric on the spindle is given by
	\begin{equation}
		\dd s^2_{\sspindle}=\f{w}{f(w)}\dd w^2+\f{f(w)}{4H(w)} \fr_z^2\dd z^2\, .
	\end{equation}
	We want to take various limits to recover the constant curvature metrics:\footnote{In the case of $\kappa=-1$ one may also consider the case where $B(x)=x^2-1$. The difference is between considering global coordinates or Poincare patch coordinates for the hyperbolic metric. Note that all that matters in the function $B$ are the signs, one could add arbitrary constants however they may all be removed by rescaling up to their sign.}
	\begin{equation}
		\dd s^2_{\Sigma_g}=\frac{\dd x^2}{B(x)}+B(x)\dd\phi^2\, ,\qquad B(x)=1-\kappa x^2\, .
	\end{equation}
	To do so it turns out to be useful to first make the following redefinitions,
	\begin{equation}
		w\rightarrow w_0 +\epsilon \gamma(x+\beta)\, , \quad z\rightarrow \frac{\phi}{\epsilon}\, ,\quad q_{i}\rightarrow r_i+\epsilon s_i\, ,
	\end{equation}
	with $\beta,\gamma, s_i, r_i$ constants and $\epsilon$ a parameter we will eventually send to 0. Next, in order to guarantee a well-defined $\epsilon\rightarrow 0$ limit we need to suitably tune the various parameters. First of all, we require the function $f(w)$ to have a double root, which fixes two of the constants to be
	\begin{equation}
		r_1=\frac{w_0^2(w_0(4-w_0)-r_2)}{r_2+w_0^2}\, ,\quad s_1=\frac{2\beta \left[ w_0\left(w_0^3(2-w_0)+2 r_2w_0(3-w_0)-r_2^2\right) - 2 s_2 w_0^3 \right]}{(r_2+w_0^2)^2}\, .
	\end{equation}
	while simultaneously fixing a third constant to be one of the branches of the equation,
	\begin{equation}
		r_2= 3 w_0-w_0^2\pm \sqrt{9 w_0^2 -4 w_0^3}\, .
	\end{equation}
	Note that we are being careful with not expanding the square root term so that we do not need to consider a further branching of solutions for whether $w_0$ is positive or not.\footnote{This turns out to be overkill, since regularity only allows for $w_0>0$. However let us continue here without using hindsight for the moment.} With these conditions at hand we end up with $f(w)$ having a double root in the limit of $\epsilon\rightarrow 0$. We will refer to the two branches as the positive branch and negative branch depending on the sign in the expression of $r_2$. Note that most parameters can be removed by simple coordinate redefinitions, so that in the end we are naively left with a two-parameter family of solutions, parameterised by $w_0$ and $s_2$. However, further analysis shows that $s_2$ is not a physical parameter, leading to a single parameter family of solutions. Furthermore, in order for the solution to be physical and eventually reduce to the BBBW setup, we must require that the scalar fields are positive definite constants. We find:
	\begin{equation}
		X_1=\frac{r_2+w_0^2}{2^{6/5} w_0^{7/5}}\, ,\quad X_2=\frac{2^{4/5}w_0^{8/5}}{r_2+w_0^2}\, ,
	\end{equation}
	which are positive definite for both branches of solutions if and only if
	\begin{equation}
		0<w_0<\frac{9}{4}\, .
	\end{equation}
	This expression for the scalars makes it clear that the two branches are in fact related by the interchange of the $1$ and $2$ indices. We may therefore restrict to studying only a single branch, which we choose to be the positive branch. 
	
	We divide the remaining steps in three cases, depending on whether $0<w_0<\tfrac{3}{2}$, $w_0=\tfrac{3}{2}$ or $\tfrac{3}{2}<w_0<\tfrac{9}{4}$. For these three cases we recover the solutions with a smooth Riemann surface of genus $g=0$, genus $g=1$ or genus $g>1$ respectively. Let us now consider these three cases separately in more detail.
	
	\subsection{Two-sphere}         %
	
	Let us start with the range $0<w_0<\tfrac{3}{2}$. We parameterise $w_0$ as
	\begin{equation}
		w_0=\frac{3}{2}\cos^2\theta_0\, ,
	\end{equation}
	and fix the free (redundant) parameters $\gamma, \delta$ to be
	\begin{equation}
		\gamma=\frac{s_2}{3\sin^2\theta_0\sqrt{6-2\cos^2\theta_0+2\sqrt{9-6 \cos^2\theta_0}}}\, ,\quad \delta=\frac{\cos^2\theta_0}{\gamma \fr_z \sin^2\theta_0}\, .
	\end{equation}
	This choice then leads to the metric 
	\begin{equation}
		\dd s^2_{\sspindle}= \frac{1}{6\sin^2\theta_0}\dd s^2(S^2)\, .
	\end{equation}
	We can then identify the parameter $z$ of the BBBW solution as,
	\begin{equation}
		z=\frac{\sqrt{3-2\cos^2\theta_0}}{\sqrt{3}\sin^2\theta_0}\, ,
	\end{equation}
	which indeed satisfies the $z>1$ constraint and indeed, the whole solution reduces to the $\AdS_5\times S^2$ solution presented in \cite{Bah:2011vv,Bah:2012dg}.
	
	\subsection{Torus}          %
	
	Next, the torus can be obtained by fixing the constants 
	\begin{equation}
		\gamma=\frac{s_2}{\sqrt{2+\sqrt{3}}\sqrt{3-2w_0}}\, ,\quad \delta=\frac{6}{\fr_z s_2}\sqrt{\frac{3-2 w_0}{2-\sqrt{3}}}\, .
	\end{equation}
	Taking the $w_0 \rightarrow \frac{3}{2}$ limit we precisely recover the $\AdS_5\times T^2$ solution of \cite{Bah:2011vv,Bah:2012dg}.
	
	\subsection{Higher genus Riemann surface}       %
	
	Finally, in order to obtain the BBBW solution with higher genus $g>1$ Riemann surface we need to take $\tfrac{3}{2}<w_0<\tfrac{9}{4}$. Similar to the two-sphere case, we parameterise $w_0$ as,
	\begin{equation}
		w_0=\frac{3}{4}( \cos^2\theta_0+2)\, ,
	\end{equation}
	and fix the redundant parameters to be
	\begin{equation}
		\gamma=\frac{2s_2\sqrt{4-\cos^2\theta_0-2\sqrt{3}\sin\theta_0}}{3\cos^2\theta_0(2+\cos^2\theta_0)}\, ,\quad \delta=\frac{(2+\cos^2\theta_0)}{\fr_z \cos^2\theta_0\gamma}\, .
	\end{equation}
	In order to complete the comparison with the solutions in BBBW we identify the parameter $z$ with,
	\begin{equation}
		z=\frac{2\sin\theta_0}{\sqrt{3}\cos^2\theta_0}\, .
	\end{equation}
	Again we see that the maps to the correct parameter range for $z$, $z>0$, and we recover the full BBBW solution \cite{Bah:2011vv,Bah:2012dg}. 
	
	Thus we have successfully shown that the local solution from which the spindle is a global completion contains all the BBBW solutions as particular scaling limits. 
	
	\newpage
	\bibliographystyle{JHEP}
	\bibliography{spindleBib}		
\end{document}